\documentclass[journal=langmuir,manuscript=article]{achemso}
\usepackage{url} 
\usepackage{graphicx}
\usepackage{amsmath}
\usepackage{lineno}
\usepackage{soul}

\title{Limits to crystallization pressure}

\author{Lei Li}
\altaffiliation{Present address: College of Physics and Optoelectronic Engineering, Shenzhen University, Shenzhen 518060, China}
\author{Felix Kohler}
\author{Joanna Dziadkowiec}
\author{Anja R{\o}yne} 
\affiliation{Physics of Geological Processes (PGP), The NJORD Centre, Department of Physics, University of Oslo, PObox 1048 Blindern, 0316 Oslo}
\author{Rosa M. Espinosa Marzal}
\affiliation{Environmental Engineering and Science, Department of Civil and Environmental Engineering, University of Illinois at Urbana–Champaign, Urbana, IL 61801}
\author{Fernando Bresme}
\affiliation{Department of Chemistry, Molecular Sciences Research Hub Imperial College, W12 0BZ, London, United Kingdom}
\author{Espen Jettestuen}
\affiliation{Norce Research, Essendropsgate 3, 0368 Oslo, Norway}
\affiliation{Physics of Geological Processes (PGP), The NJORD Centre, Department of Physics, University of Oslo, PObox 1048 Blindern, 0316 Oslo}
\author{Dag Kristian Dysthe}
\affiliation{Physics of Geological Processes (PGP), The NJORD Centre, Department of Physics, University of Oslo, PObox 1048 Blindern, 0316 Oslo}
\email{d.k.dysthe@fys.uio.no}

\begin{document}

%
%
\begin{abstract}
\begin{center}
\includegraphics[width=8.2cm]{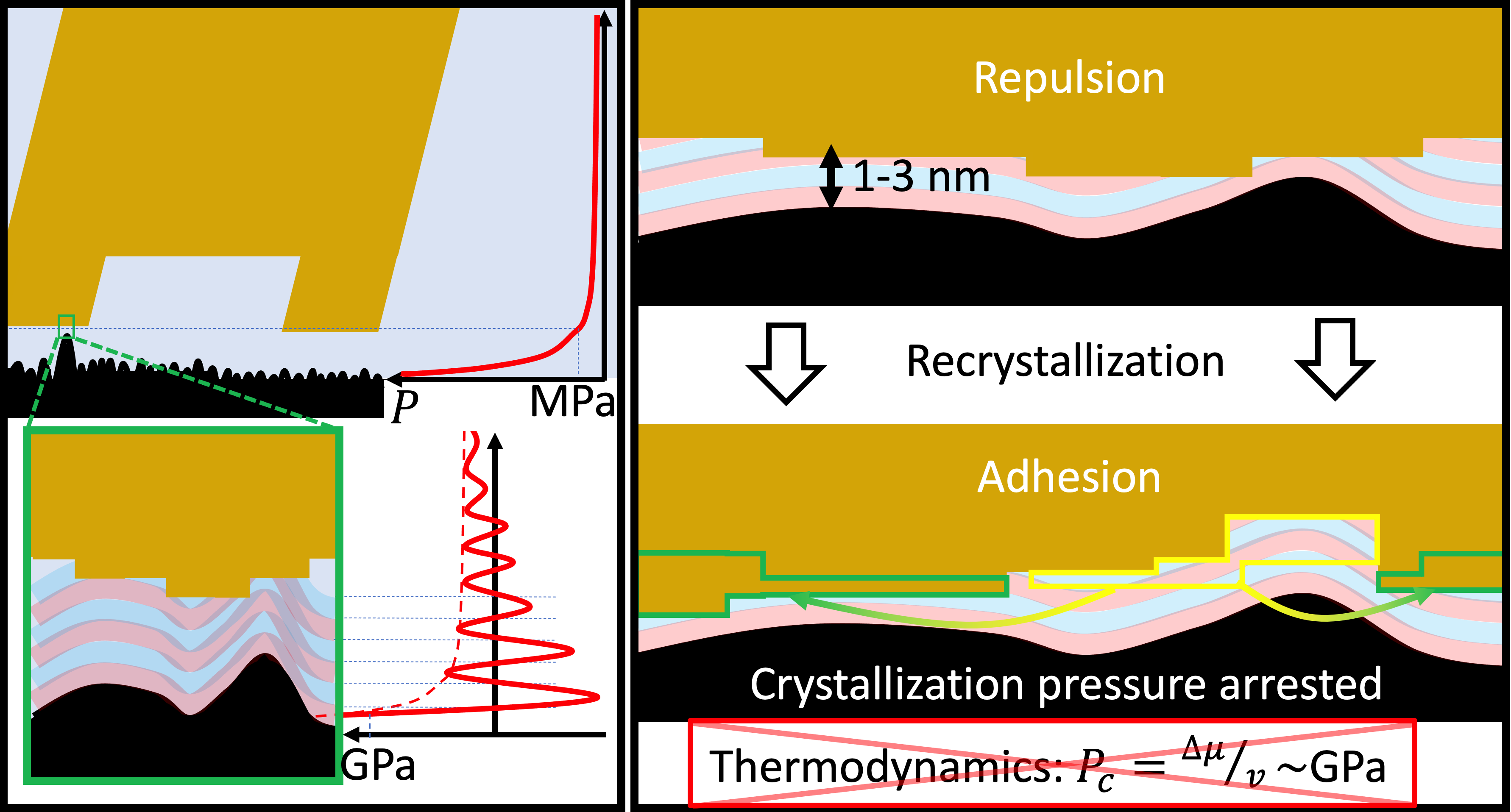}
\end{center}

Crystallization pressure drives deformation and damage in monuments, buildings and the Earth's crust. Even though the phenomenon has been known for 170 years there is no agreement between theoretical calculations of the maximum attainable pressure and that found experimentally. We have therefore developed a novel experimental technique to image the nano-confined crystallization process while controlling the pressure and applied it to calcite. The results show that displacement by crystallization pressure is arrested at pressures well below the thermodynamic limit. We use existing molecular  dynamics simulations and atomic force microscopy data to construct a robust model of the disjoining pressure in this system and thereby calculate the absolute distance between the surfaces. Based on the high resolution experiments and modelling we formulate a novel mechanism for the transition between damage and adhesion by crystallization that may find application in Earth and materials sciences and in conservation of cultural heritage.
\end{abstract}

\maketitle

\section{Introduction}
Crystallization pressure is well known to induce fracture and deformation in solids confining crystals, damaging buildings and monuments~\cite{Espinosa-Marzal2010,Flatt2014a}, lifting layers of the Earth's surface~\cite{Gratier2012a} and it is thought to drive vein formation~\cite{Taber1928,Wiltschko2001}, spheroidal weathering~\cite{Royne2008} and cracking during metamorphism and frictional failure of the Earth's crust~\cite{Kelemen2012}. 

The question of what limits the crystallization pressure is important to mitigation, repair and conservation of buildings and monuments damaged by ``salt crystallization''. Different treatments in stone conservation aim at altering the surface energy of pore surfaces in order to limit water transport, controlling the regions where crystallization occurs and by limiting the crystallization pressure itself~\cite{Rodriguez-Navarro2000,Espinosa-Marzal2010,Desarnaud2016,Jia2019}. In Earth sciences it is fundamentally important to know if a weathering reaction or a metamorphic reaction may generate a crystallization pressure sufficient to fracture the surrounding rock, opening new fluid pathways for further reaction and frictional failure. Recent estimates for olivine hydration and carbonation suggest pressures of the order of 1~GPa can be reached~\cite{Kelemen2012}, whereas recent experiments show that the fracture process driven by crystallization pressure closes down long before such a pressure limit is reached~\cite{Zheng2018}. 

The crystallization pressure is generated by a crystal growing in a "load bearing" grain boundary/contact area. A load bearing grain boundary/contact area is an area where the solid grains transmit stress to each other 1) either through direct interatomic interactions between the solids -- a {\em solid-solid contact} 2) or transmitted through a thin ($<50$~nm) layer of fluid, where the pressure supported by the fluid is called the {\em disjoining pressure}. In the first case, the solid nature of the grain boundary inhibits mass transport except at high temperature. In the second case, mass may be transported in the fluid layer to the growing crystal. The driving force of the mass transport, crystal growth and crystallization pressure is the supersaturation of the fluid present. The existence of a crystallization pressure has been observed and demonstrated many times during the last 170 years~\cite{Becker1905,Becker1916,Taber1928,Flatt2007,Sekine2011,Royne2012b,Naillon2018,Desarnaud2016,Kohler2018,Li2018} and the thermodynamic limit to this pressure, $P_c$ has been known since the work of Correns and Steinborn~\cite{Flatt2007}: $P_c=\Delta\mu/v$, where $\Delta\mu$ is chemical potential of the solution relative to the equilibrium state and $v$ is the molar volume of the crystal. Apart from the somewhat dubious results of Correns~\cite{Flatt2007}, no-one has ever reported crystallization pressures approaching the thermodynamic limit~\cite{Flatt2007,Sekine2011,Royne2012b,Desarnaud2016}. There are three main candidates to explain the discrepancy: 1) As already observed in 1915, the load bearing contact area is much smaller than the apparent contact area~\cite{Becker1916,Weyl1959,Royne2012b,Kohler2018,Li2018}, meaning that the pressure in the load bearing contact area may possibly be approaching the thermodynamic limit. 2) Due to mass transport by diffusion, the supersaturation in the contact is smaller than in the bulk solution~\cite{Sekine2011,Naillon2018,Kohler2018}. 3) The fluid film in the contact ``collapses'', a stable, "close contact" is created and diffusion mass transport and crystal growth stops before the thermodynamic limit of crystallisation pressure is reached.

In order to understand what limits the disjoining pressure, we develop a new experimental setup where we control and measure the exact supersaturation, the real load bearing contact area, the crystal growth rate and we use existing simulation results and AFM data to construct a model of the disjoining pressure and diffusion of the system that we study experimentally. We conclude that the equilibrium concepts of crystallization pressure and disjoining pressure are not sufficient to explain the experimental results. Hence, we propose a new mechanism: The reactive surface grows locally to increase the adhesive surface area and the adhesive energy between the solid surfaces separated by 2-3 water layers becomes larger than the energy associated to the supersaturation driving the crystallization (see Figure \ref{fig:mechanism}).

\begin{figure}[thb]
\includegraphics[width=8cm]{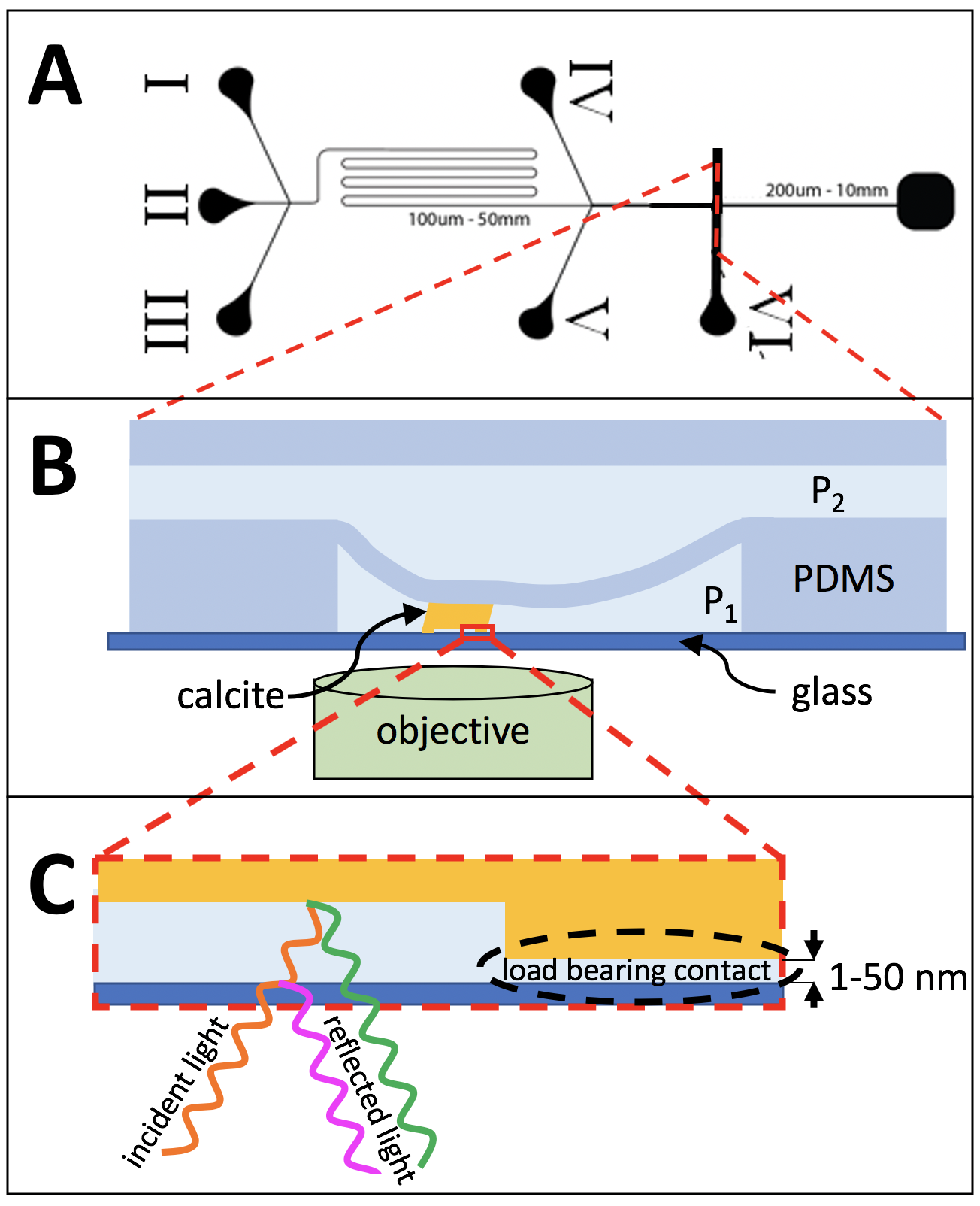}
\caption{{\bf Microfluidic experiment with growth control, pressure control and in situ interferometric (RICM) imaging.} {\bf A:} Flow control pattern with five inlets I-V to vary the CaCO$_3$ concentration and thereby control nucleation and growth of calcite in the channel (as already published in~\citep{Li2018}). {\bf B:} Two-layer PDMS on glass microfluidic channels for flow and concentration control in lower channel (pressure $P_1$) via inlets I-V and, novel in this study: control of force between calcite and glass by $P_2>P_1$ controlled via inlet VI. {\bf C:} In situ imaging with height measurement by interference between light reflected from glass-fluid interface (pink) and fluid-calcite interface (green). This allows identification of load bearing  contacts (glass-calcite distance $h<50$~nm) and crystal growth rate in these contacts.}
\label{fig:calcitesetup}
\end{figure}

\section{Experimental}
\subsection*{Microfluidic device with pressure control channel}
The microfluidic device, which is shown in Figure~\ref{fig:calcitesetup}, consists of a cover glass with two PDMS layers on top. The PDMS is attached to the glass and defines two layers of fluid channels. The lower layer is used to control nucleation and growth of calcite. Calcite nucleation and dissolution of nuclei is repeated until a nucleus is attached on the PDMS membrane in the desired region. The fluid pressure in the upper fluid layer can be increased in order to bend the 6~$\mu$m thick PDMS membrane between the two fluid layers and push the calcite crystal towards the glass. In this manner the force between the calcite crystal and the glass can be controlled. The microscope objective underneath the cover glass images the crystal and allows full 2D measurement of the absolute vertical distance between the crystal surface and the glass surface. The cover glass has an RMS roughness of 0.2~nm as determined by AFM measurements. The force divided by load bearing contact area yields the contact pressures and the vertical displacements yields the crystal growth rates in the load bearing contacts. 

The layout of the lower layer corresponds to the one described previously~\cite{Li2017b}. However, it is only 29 $\pm$ 0.3 $\mu$m deep. It has 5 inlets (I-V), which are used to control the nucleation and growth of calcite in the fluid flow. The CaCl$_2$, H$_2$O and Na$_2$CO$_3$ solutions, which are pumped into the microfluidic cell from inlets I,II and III, mix in the main channel by diffusion. To induce nucleation, we use solutions with higher concentrations of CaCl$_2$ and Na$_2$CO$_3$ from inlets IV and V. The highly concentrated CaCO$_3$ solution produces nuclei that are attached to the walls of the channel. Multiple nucleations or nuclei at undesired locations are dissolved by lowering the concentration of the solution. After nucleation, a CaCO$_3$ concentration of 0.801$\pm$0.002~mM has been used, which corresponds to a saturation index of $\Omega=0.44$~\cite{Li2017b}.

Before the calcite crystal comes into contact with the glass bottom of the microfluidic channel the pressure $P_2-P_1$ deforms the membrane. Once a load bearing contact is achieved the crystal moves negligibly and the force on the crystal from the membrane is proportional to the change in pressure $P_2$ times the area, $A_m$, of the membrane closest to the crystal, $F=\Delta P_2 A$ (see section~\ref{sec:force}).

\subsection*{Reflection interference measurement of fluid film thickness}
The images of the confined crystal interface have a local intensity $I$ that depends on the fluid film thickness or local distance $h$ between the glass surface and the crystal: $I=I_0+I_1cos(4\pi hn\alpha_\theta/\lambda + \pi)$, where $n=1.33$ is the refractive index of water, $\lambda=550$~nm is the wavelength of our light source, $I_0$ is the background intensity and $I_1/I_0$ is the contrast and $\alpha_\theta\approx 1$ is a factor that accounts for the effective angle of the light with respect to the optical axis. Because $I_0$ varies across the image due to refractions and non-uniform reflections at other surfaces of the crystal, there is an uncertainty of about $\pm 10$~nm in the determination of contact ($h=0$). The accuracy of the determination of upwards growth $\Delta z=z(t+\Delta t)-z(t)$ (where $z$ is the vertical position  of the crystal) equals the precision of $\pm 0.5$~nm. The details of reflection interference contrast microscopy (RICM) have been explained in detail elsewhere~\cite{Li2018,Kohler2018}. From the RICM images we identify the load bearing contacts as regions where the distance between the crystal and the glass is less than 50~nm.

\subsection*{Thermodynamics and kinetics of calcite growth}
The saturation index $\Omega$ is related to the chemical potential of the solution:
\begin{equation}
\Omega=\frac{\Delta\mu}{kT}=\ln\left(\frac{a_{Ca^{2+}}a_{CO_3^{2-}} }{K_{sp}}\right),
\end{equation}
where $a_{Ca^{2+}}$ and $a_{CO_3^{2-}}$ are the ion activities. Teng et al.~\cite{Teng2000} have proposed that the solubility product $K_{sp}=10^{-8.54}$ corresponds to the experimental conditions when
spirals on the 10$\bar{1}$4 surface stopped growing.  We have used PHREEQC~\cite{Charlton2011} to calculate $\Omega$.

A normal stress, or (load bearing contact) pressure, $P_c$ on a solid surface contributes with a factor $P_cv$ to the chemical potential of the solid, where $v$ is the molecular volume of the solid~\cite{Dysthe2014}. Thus the chemical potential difference between the solid and the solution that drives either growth ($\Delta\mu>0$) or dissolution ($\Delta\mu<0$) is:
\begin{equation}
  \Delta\mu=kT\Omega - P_cv.
  \label{eq:deltamu}
\end{equation}
One may then immediately calculate that a solution with saturation index of 0.44 is in equilibrium with a calcite surface subject to a pressure of $P_c=35$~MPa. This is the thermodynamic limit of the crystallization pressure at saturation index 0.44. We may also calculate that the disjoining pressure of 0.5-5~MPa in the contacts amounts to reducing the driving force for growth, $\Delta\mu/kT$ by 1.4-14\% from 0.44 to 0.43 and 0.38, respectively.  We have shown that in this range of saturation indices the purely kinetic growth rate constant (no diffusion limitation) is independent of saturation index~\cite{Li2018} and the kinetic contribution to the growth rate will therefore only be reduced by 1.4-14{\%}.

\section{Results}
\subsection*{Experimental results}
The novel experiments presented here have been designed for \textit{in situ} observations of nano-confined calcite growth under highly controlled conditions. The microfluidic setup provides a very accurate and stable supersaturation and has a high degree of control of the pressure at the confined surface. The topography of the nano-confined calcite surface and thereby the load bearing contacts (glass-calcite distance $h<50$~nm) are recorded with nanometer vertical resolution during the whole growth process by high resolution reflection interference contrast microscopy (RICM).

\begin{figure}[th]
\includegraphics[width=9cm]{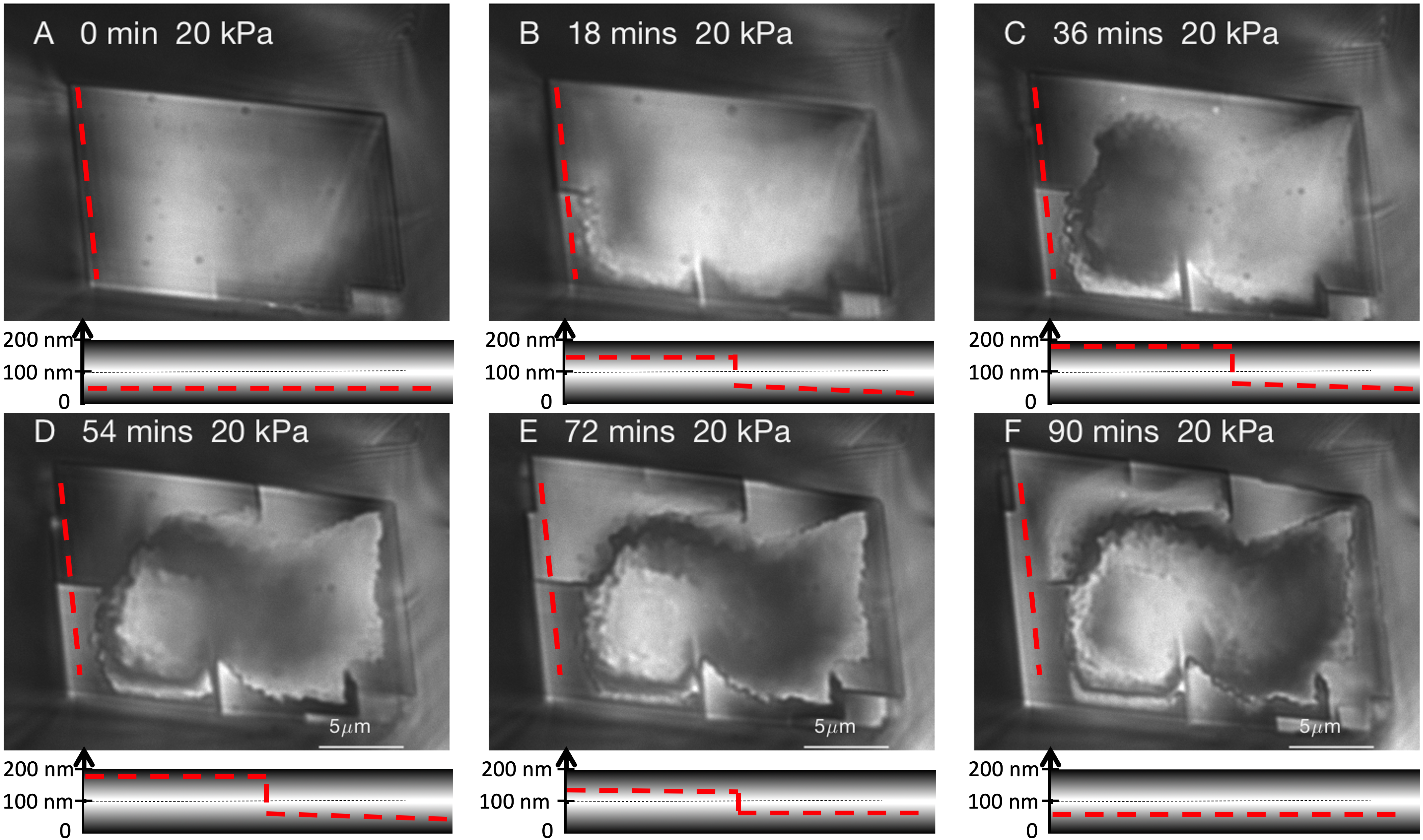}
\includegraphics[width=9cm]{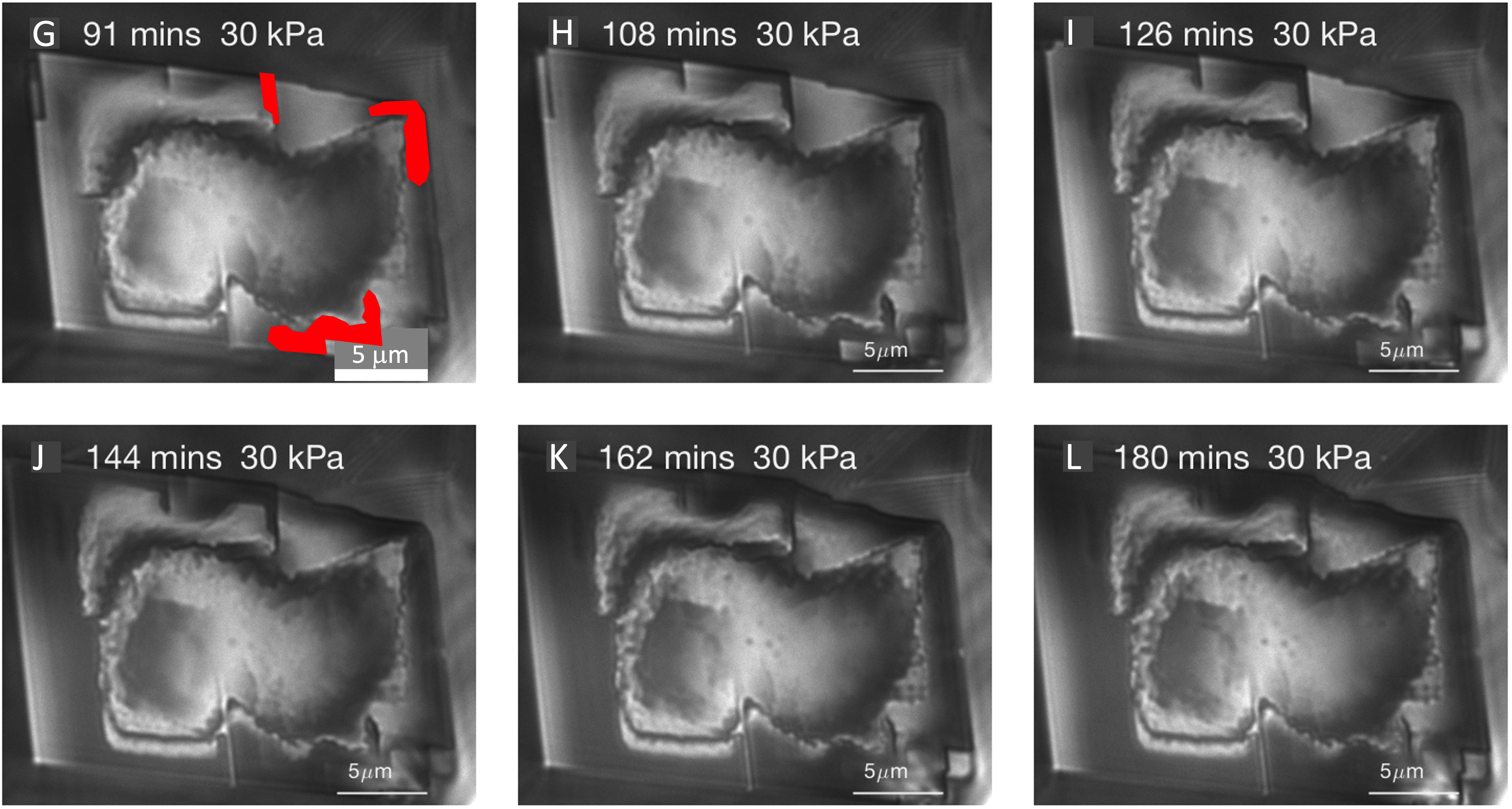}
\caption{{\bf Nano-confined calcite surface evolution.} \small RICM images of the nano-confined crystal growth of calcite at low load (top 6 images) and high load (bottom 6 images). The intensity, $I$, in the images indicate height, $h$, above the glass surface: $I\propto -\cos(2\pi h/\mathrm{207 nm})$. The height profile along the red line at the left side of the crystal is indicated in the first 6 images. The height information is used to identify load bearing contact area as indicated for image G. {\bf Images A-F}: Evolution from $t=0$ when saturation index of the solution is increased from 0 to 0.44, applied pressure is $P_2$=20~kPa and the force transmitted to the crystal is $F\approx 7$~nN. At this low load the load bearing contacts ($h<50$~nm) change dynamically. {\bf Images G-L}: Evolution from $t=91$~min when applied pressure is increased to $P_2$=30~kPa and the force transmitted to the crystal is $F$=64~$\mu$N. At this high load the load bearing contacts merge and eventually cover the entire rim. A timelapse movie of the experiment is available~\cite{SupInf}.}
\label{fig:calcitepress}
\end{figure}

\begin{figure}[th]
\includegraphics[width=9cm]{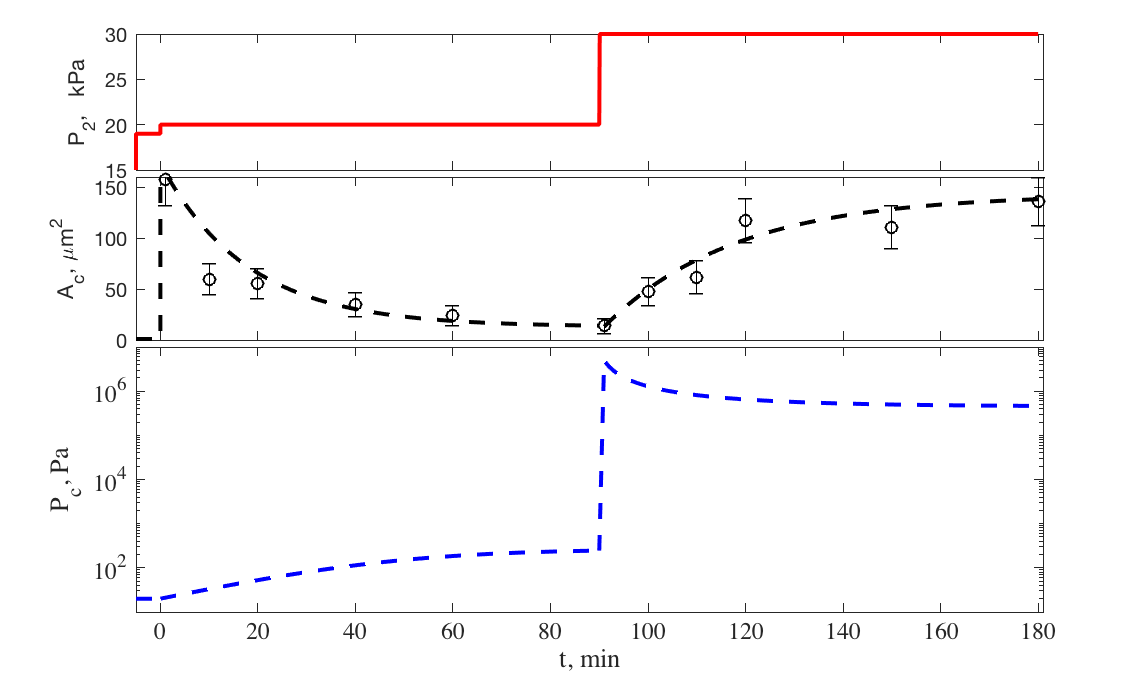}
\caption{{\bf Control fluid pressure, contact areas and contact pressures.} {\bf Top:} Imposed control fluid pressure as function of time. {\bf Middle:} Load bearing contact area $A_c$ (circles with error bars) is the area of crystal-glass distance smaller than 50~nm as measured in  the interferometry images (see Figure~\ref{fig:calcitepress}) and smooth fit of $A_c$ (dashed line). {\bf Bottom:} Pressure in load bearing contacts between calcite and glass support as function of time.}
\label{fig:contactareas}
\end{figure}

We have succeed to nucleate and grow calcite crystals attached to the deformable membrane in several experiments. We have proceeded by increasing the control pressure to form a load bearing calcite-glass contact. All experiments have given qualitatively the same results but we focus here on the experiment where the contact stresses could be determined quantitatively and thereby be analysed properly.

While bringing the crystal into contact with the glass, the flowing calcium carbonate concentration is kept at $c_{CaCO_3}$=0.05~mM and the saturation index $\Omega$=0. The crystal is brought into contact with the glass surface with a fluid control pressure $P_2$=20~kPa and the force transmitted to the upper calcite surface is $F\approx 7$~nN (see section~\ref{sec:force}). The flowing fluid composition is changed to $c_{CaCO_3}$=0.8~mM, $\Omega$=0.44 at time $t=0$. Figure~\ref{fig:calcitepress} shows the evolution of the confined crystal surface after it has been brought in contact with the glass surface. The crystal grows outwards, changing the area $A$ of the crystal parallel to the glass. The crystal also grows downwards in the load bearing contacts (perpendicular to the glass surface and image plane) pushing the crystal upwards against gravity and the applied force of the membrane.

\begin{figure}[th]
\includegraphics[width=9cm]{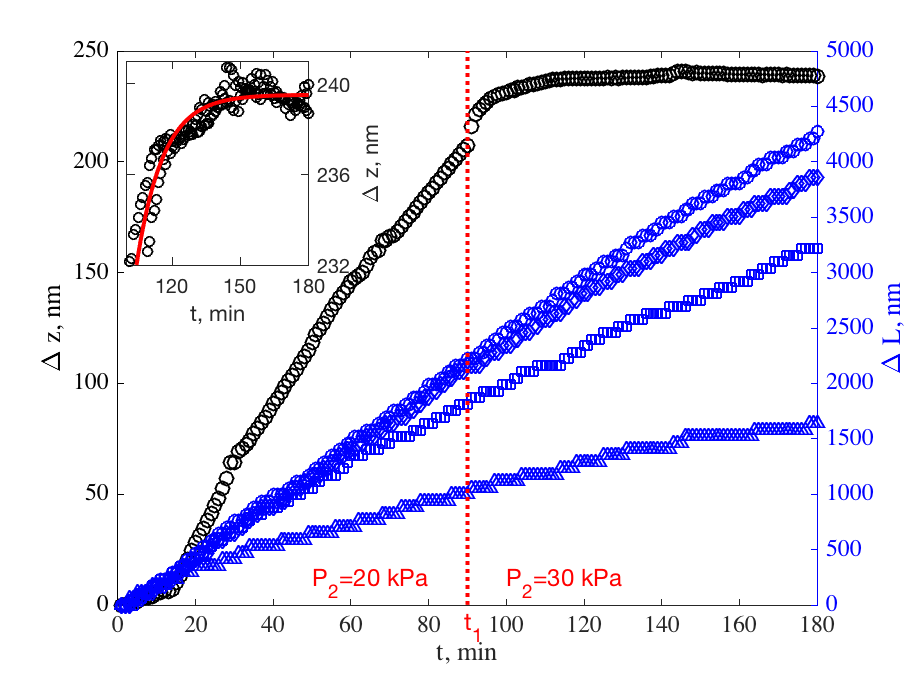}
\caption{{\bf Calcite growth rate} in all directions before and after fluid pressure, $P_2$, increase at $t_1=90$~min. Blue curves: lateral growth, no effect on rate of changing fluid pressure $P_2$. Black data points: vertical growth comes to halt when pressure is increased. Inset: exponential decay of rate after pressure increase measured at two different areas in the middle of the crystal. The standard deviation of the height measurement during the last 40 minutes is 0.5~nm.}
\label{fig:calcitesize}
\end{figure}

At time $t=0$ the calcite surface was flat ($h=40\pm 10$~nm, Figure~\ref{fig:calcitepress}~A, top) and glass-calcite contact pressure $P_c=20\pm 10$~Pa. The calcite surface was confined and the Ca$^{2+}$ and CO$_3^{2-}$ diffusion is limited in the confined solution film causing the first contact growth rim to form (see Figure~\ref{fig:calcitepress}~B, top). As the calcite grows at the confined surface, a cavity appears on the calcite surface, a confined growth transition that we have already explained in detail~\cite{Li2017b,Kohler2018}. Along the rim of the crystal the crystal grows and the dark part of the growth rim signifies a small distance $h<50$~nm and that this area, $A_c$, of the crystal is load bearing. One observes in images C-F that less than 5\% of the crystal area is load bearing. The growth rim is divided into different domains separated by large steps. It changes with time which domains of the rim that are load bearing. This causes the crystal to "wobble upwards" in a manner already reported for crystals only supporting their own weight~\cite{Li2017b}. 

We have used the RICM images in Figures~\ref{fig:calcitepress} to estimate the contact areas, $A_c$ between the calcite and the glass support (see Figure~\ref{fig:contactareas}). The contact pressure is then calculated as $P_c=(P_2-P_{2,0})A_m/A_c$, where $P_{2,0}=19.26$~kPa is the pressure used to stretch the membrane until full contact between calcite and glass was reached and $A_m$ is the area of the membrane transferring pressure from the control fluid to the crystal (see section~\ref{sec:force}). In the first 90 minutes the contact pressure increased from 20 to 250~Pa, at $t=90$~min the contact pressure changed from 250~Pa to 5~MPa and then the contact pressure reduced towards 0.5~MPa as the contact area grew.

After 1.5 hours growth with saturation index $\Omega =$0.44, the pressure in the upper channel is increased from $P_2=20$~kPa to 30~kPa and the saturation index is kept constant. The images of the crystal in Figure~\ref{fig:calcitepress}~G-L show that the growth rim that was split up in domains with steps between, grows to form a smooth calcite rim in contact with the glass surface all around the crystal rim. During the first 30 minutes of this reformation of the rim the crystal growth is still pushing the crystal up against the applied load (see Figure~\ref{fig:calcitesize}). After that, the upwards growth stops and the contacting rim widens as the outer edges of the crystal continue to grow.

During the three hours (180 minutes) with high saturation index, $\Omega$=0.44, the outer rims of the calcite crystal grew at a constant rate as can be seen in Figure~\ref{fig:calcitesize}. The upwards growth shown in the same figure however, goes through three distinct phases: The first 13 minutes the confined surface grows to accommodate the contact and tilting the crystal slightly, then during the period 13-90 minutes there is a steady upwards growth of 2.6~nm/min. At $t_1$=90~min the pressure increase pushes the crystal 8~nm downwards and then the vertical growth slows down exponentially and comes to a complete halt. We continued to let the crystal grow under this load for 12 hours more, but there was no further growth upwards within the accuracy  (1 standard deviation during the last 40 minutes, see inset in Figure~\ref{fig:calcitesize}) of $\pm 0.5$~nm. We can therefore conclude that the growth rate is smaller than 0.04~nm/h which equals 0.35~$\mu$m/yrs or 35~cm/Ma. Relative to the height of the crystal this growth rate corresponds to a strain rate of $10^{-9}$~s$^{-1}$. Thus even though we are below the detection limit of a very accurate technique the growth rate and strain rate is still considerable on a geological time scale.
This result requires us to pose the question: does the change in growth rate of more than a factor 4000 signify a dramatic slowing down of the crystal growth (working against a force) or does it signify a complete stop?

\begin{figure*}[th]
\includegraphics[width=17cm]{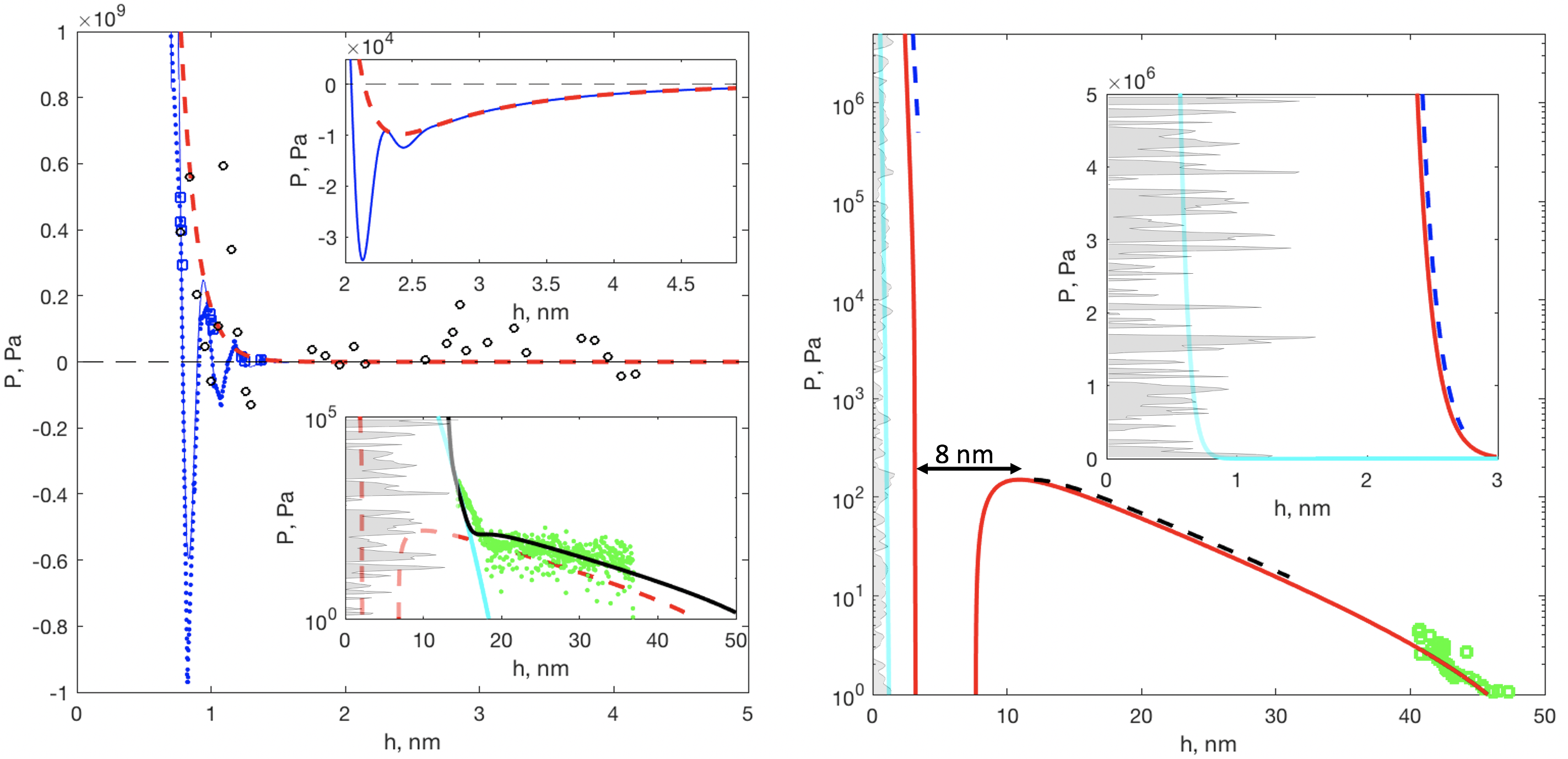}
\caption{{\bf Disjoining pressure} data and models for pure water in calcite and glass interfaces.{\bf Left:} Simulation data (blue dots and squares~\cite{Brekke-Svaland2018}, black circles~\cite{Dysthe2002}) for flat calcite surfaces and models fitted to the data (blue line: full fit, red dashed line: fit to repulsive part of MD data). {\bf Left, upper inset:} Small amplitude, long range attractive well from DLVO theory outside range of recent MD data~\cite{Brekke-Svaland2018}. {\bf Left, lower inset:} Green dots: experimental data for rough, glassy silica surfaces on atomically flat calcite surfaces~\cite{Diao2016}. Black line: Parsons model with $\sigma=3$~nm, cyan line: contact part of Parson model, gray: illustration of roughness with  $\sigma=3$~nm.
  {\bf Right:} Disjoining pressure model for load bearing contacts in this experiment, atomically flat calcite on rough glass with  $\sigma=0.2$~nm (red line). Green squares: experimental data from~\cite{Li2017b}. Cyan line: contact part of Parson model, gray: illustration of roughness with  $\sigma=0.2$~nm. Black dashed line: Pressure -- distance range of this experiment before pressure increase at 90 min. Blue dashed lines: Pressure -- distance range of this experiment after pressure increase at 90 min. As indicated, increasing the pressure above 250~MPa causes an 8~nm jump in $h$.}\label{fig:disjoining}
\end{figure*}

\subsection*{Model results}
In this section we combine experimental data and simulation data, with a theoretical model that accounts for the forces between rough surfaces, to build a new model to describe the disjoining pressure between calcite and a rough glass surface. Interaction energies between surfaces are represented  with DLVO forces at distances of a few nanometers and beyond, and with repulsive and oscillatory forces at smaller distances~\cite{Israelachvili2011}.

We have fitted a simple repulsive model, $P_{SH}=5\cdot 10^{11}\exp(-1.3h/\lambda)$ and oscillatory amplitude, $\Pi_O=\cos(2\pi h/(\lambda+0.03h)))$ to the Molecular Dynamic (MD) simulation data for calcite-calcite slit pores in pure water~\cite{Brekke-Svaland2018} (see left hand side of Figure~\ref{fig:disjoining}). Given the lack of data for calcite-glass surfaces, we will use this as the best starting point to model interactions between atomically flat calcite-glass surfaces. The period of the oscillatory force is due to the water layer structure and it does not depend much on the surface type. To take into account the charges of the calcite and silica surfaces we use the calcite-silica parameters for the DLVO model from Diao and Espinosa-Marzal\cite{Diao2016}. The inset on the left side of Figure~\ref{fig:disjoining} shows the AFM data for rough silica spheres on calcite~\cite{Diao2016}, featuring a steep repulsive part $P>100$~Pa and a long range tail of $P$=10-100~Pa. The long range part is outside the range of the MD data and the repulsive model. To account for the steep repulsive part of the data we use the rough surface model of Parsons et al~\cite{Parsons2014} with an RMS roughness of 3~nm. The resulting disjoining pressure model can be expressed as $P(h,\sigma)=(P_{SH}(h)+P_{DLVO}(h))\otimes p_G(h,\sigma)$, where $\otimes$ represents the convolution with the Gaussian height distribution $p_G(h,\sigma)$ with $\sigma$ the RMS roughness. The only adjustable parameter required to obtain the black curve that passes through the AFM data, was the RMS roughness, $\sigma$, of the silica sphere. We used here $\sigma$=3~nm, close to the value reported by the authors, $\sigma$=2~nm, in reference~\cite{Diao2016}. One observes that the reason why no attractive interaction was observed between the silica sphere and the calcite is that the direct contact contribution to the pressure between the two surfaces (cyan line in Figure~\ref{fig:disjoining}) extends further out than the negative DVLO contribution (dashed red line in Figure~\ref{fig:disjoining}).

The same model is then used to predict the disjoining pressure curve for our experiments with glass roughness $\sigma=0.2$~nm (see right hand side of Figure~\ref{fig:disjoining}). The model fits very well with our previously published data (green squares). The model predicts the range of distances between the two surfaces in the first 90 minutes (10-250~Pa $\Rightarrow$ 10-30~nm) and in the last 90 minutes (5$\cdot 10^5$-5$\cdot 10^6$~Pa $\Rightarrow$ 2.4-2.6~nm). This corresponds well with the observed jump of $8\pm 2$~nm when the pressure was increased. In Figure~\ref{fig:disjoining} one also observes that there are no solid-solid contacts between the two surfaces (see gray lines in Figure 6, which represent a 0.2~nm RMS rough surface, while the cyan line is the contact contribution of the Parson model).

The change in the gap, $\Delta h=8\pm 2$~nm, between the two surfaces influences the mass flow rate towards the growing surface in two ways: i) the cross sectional area for the flow is reduced by a factor 5 (from 10 to 2~nm), ii) the diffusion coefficient is reduced a factor of 2.5 (see section~\ref{sec:diffusion}). In addition the diffusion flux is reduced inversely with the width of the growth rim, $w$. Figure~\ref{fig:calcitepress} shows that during the experiment the rim width only changes by a factor of 2-4. Altogether, this amounts to at most a factor 50 reduction in the vertical growth rate. The observed growth rate is reduced by {\em at least} a factor 4000 but the reduction of the diffusion due to confinement can only account for only a factor 50. The increased pressure in the load bearing contacts affects the growth rate of calcite (through the chemical potential, see section Experimental) by 14\% at the most.
This means that some other mechanism is needed to explain a growth rate reduction of a factor 80 or more.

\section{Discussion}
\begin{figure*}[th]
\includegraphics[width=10cm]{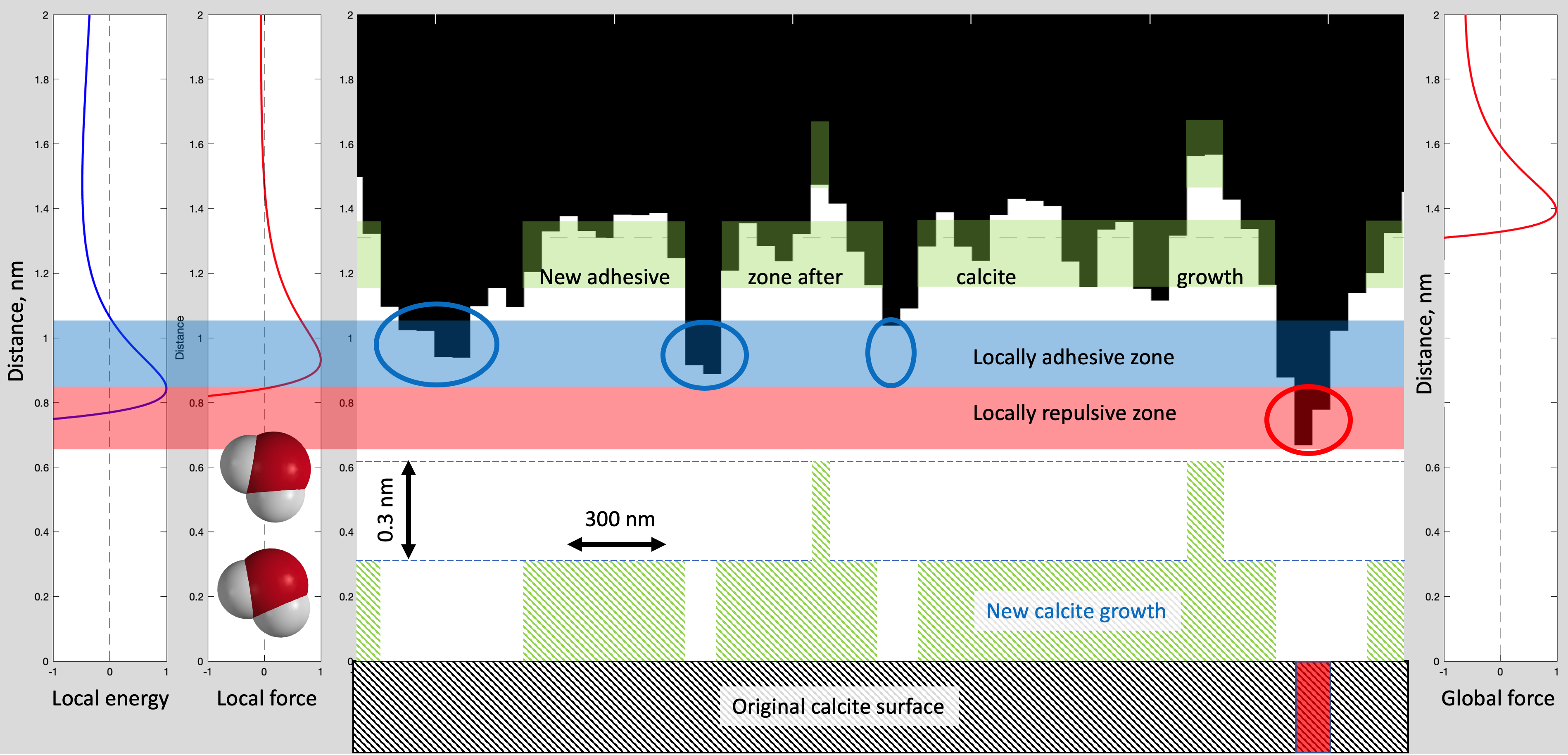}
\caption{{\bf Adhesion arrests repulsive crystal growth}
A rough glass surface (top) is pressed beyond the long range repulsive forces towards an initially flat crystal surface (bottom, in this figure the crystal is calcite). Note: the vertical scale is 1000-fold exaggerated. The crystal surface is reactive and growing due to the solution between the surfaces being supersaturated. The local energy and force curves (left hand) are integrated over the rough surface to obtain the global force curve (right hand) acting on the mean height of the roughs surface (dashed line). The protruding load bearing contacts (red circle) displaces the global force curve (right hand) with respect to the local force curve (left hand). Some parts of the surfaces are in adhesive contact (blue region). These regions on the crystal surface will grow quickly and the total adhesive force will become larger than the maximum repulsive force available by crystal growth. In addition the diffusion of water and ions goes to zero as the distance between the surfaces is reduced to 2-3 molecular layers of water (the inset water molecules are to scale). Consequently, maximum crystallization pressure is not given by supersaturation as predicted by thermodynamics, but by surface interaction parameters like interaction energy and roughness.}
\label{fig:mechanism}
\end{figure*}

We have shown that when the pressure at the confined interface is increased enough to establish close proximity ($h\approx 2.5$~nm), the confined vertical motion of the crystal is reduced by at least a factor 4000 and apparently stops completely. We have observed the same arrest of vertical motion in many other experiments~\cite{Li2017b,Kohler2018}. The reduced diffusion transport of ions to the confined surface and the change in thermodynamic driving force or growth rate kinetics may all together account for a reduction by a factor of 60 only. 

To explain the arrest of the vertical motion we propose a new mechanism. Figure~\ref{fig:mechanism}) sketches a molecular interpretation of the processes at play between the rough inert glass surface and the reactive calcite surface that we have modelled  in Figure~\ref{fig:disjoining}.
When the surfaces are pressed together by more than 1~MPa the mean distance between a rough ($\sigma$=0.2~nm) and a flat surface is 1-3~nm depending on the nature of the fluid and charges on the surfaces. However, local regions of the calcite surface are closer or more distant to the glass surface. Due to the ordering of the confined water different regions of the surfaces will experience either local adhesion or repulsion. In addition to the chemical potential $\Delta\mu$ of supersaturation there is a local free energy difference driving crystal growth in the regions where the growth results in increased adhesive energy. Local pressures exceeding $P_c=\Delta\mu/v$, will  drive local dissolution. Consequently, once the surfaces are brought into sufficiently close contact, local dissolution and growth will reshape the calcite surface to fit the glass surface, leading to a maximisation of the local regions experiencing an attractive interaction, at typical separations of 2-3 water layers. The diffusion in this water film decreases significantly, approaching zero. Consequently, crystal growth ceases and the surfaces adhere instead of being pushed apart. If the surfaces do not have appreciable adhesive regions, the calcite surface may still locally dissolve and grow approaching the other surface. This may stop diffusion and crystal growth without the surfaces adhering. The proposed mechanism will depend on the surface roughness, the hydrophilic/phobic nature of the surfaces and the fluid composition. These are interactions that can be modelled~\cite{Israelachvili2011} to predict the maximum crystallization pressure.

It has been demonstrated that liquid ordering is important during crystal growth~\cite{Freitas2020} and that crystallisation can take place in local regions. The crystallization is correlated with the observation of negative and positive disjoining pressures, which may change on very short lengthscales (nano and sub-nanometer distances)~\cite{Camara2004}. A recent study of contact formation using Kinetic Monte Carlo (KMC)~\cite{Hogberget2020a} shows that growth of local contacts is enhanced by an attractive interaction energy of the same order as that created by the ordering of 2-3 confined water layers (see section~\ref{sec:KMC}).

SFA and AFM experiments demonstrate that roughness is important for short time adhesiveness and dissolution - precipitation processes in the confined region~\cite{Royne2015,Diao2016,Dziadkowiec2018,Javadi2018,Dziadkowiec2019}. Systematic variation of contact time should allow a better understanding of the adhesion forming mechanism that we propose here. Indeed, we evidenced an analogous calcite growth mechanism in the SFA experiments with reactive calcite surfaces growing against a mica substrate (see Figure \ref{fig:SFA-mc} and details in the Supporting Information: SFA experiments). These SFA results indicate that the growing calcite asperities become locally smoother, leading to the stronger adhesion between calcite and mica with time. 

The proposed mechanism is closely related to adhesion between reactive solids and resembles the molecular scale processes proposed for crystal agglomeration~\cite{Brunsteiner2004,Brunsteiner2005}. Experiments on the interactions between reactive surfaces in the surface forces apparatus (SFA)~\cite{Dziadkowiec2018,Dziadkowiec2019}, with AFM~\cite{Royne2015,Diao2016,Javadi2018} and slide-hold-slide friction~\cite{Renard2012} all show that the adhesion between two surfaces depends on the fluid present, the force applied and time spent holding the surfaces together before pulling them apart or sliding.

The proposed mechanism is also consistent with recent experimental observations that showed that the limit to crystallization pressure is related to the disjoining pressure and not to the thermodynamic limit pressure~\cite{Desarnaud2016,Zheng2018}. Our proposed ``microfracture healing'' without forming covalent bonds, only weak water-film-mediated ``bonds'' can also explain several experimental observations of reactive interfaces developing strength with time: fracture healing~\cite{Anders2014}, cement setting~\cite{Rodriguez-Sanchez2020} and fault gouge strengthening~\cite{Renard2012,Niemeijer2008}.

Recently, it has been demonstrated that the crystallization pressure of NaCl on glass is {\em reduced exponentially} with supersaturation even though the thermodynamic limit {\em increases} with supersaturation~\cite{Desarnaud2016}. The authors argued that the crystallization process was arrested once the fluid film reached a thickness of about 1.5~nm. Our experimental and modelling study on the nanoscale explains the mechanism how crystallization pressure is arrested at fluid film thicknesses of 2-3 water layers. We also demonstrate that modelling of the surface forces including roughness may predict the limit of crystallization pressure.

A systematic evaluation of the proposed mechanism can be performed both by Kinetic Monte Carlo (KMC)~\cite{Hogberget2016,Hogberget2020a,Hogberget2020b} and experimentally combining optical imaging of the contacts with AFM experiments~\cite{Royne2015,Diao2016}. The existence and effectiveness of the proposed mechanism depends crucially on the roughness and surface forces. These are parameters that can easily be varied experimentally and in KMC. Molecular simulations of hydrated crystals like mirabilite and alum may reveal if their large crystallization pressures and damaging properties\cite{Flatt2007,Espinosa-Marzal2010,Flatt2014a} are due to qualitative differences in water structure, adhesion and diffusion as compared to non-hydrated crystals like  CaCO$_3$, NaCl~\cite{Sekine2011,Desarnaud2016,Naillon2018} NaClO$_3$~\cite{Kohler2018}.

\section{Conclusions}
A new experimental technique to control and image crystal growth in nano-confinement has been developed and applied to calcite and shown that displacement by crystallization pressure is arrested at pressures well below those corresponding to the thermodynamic limit. Existing simulation and AFM experimental data have allowed us to build a robust model to rationalize the disjoining pressure in our system and thereby calculating the absolute distance between the surfaces. Our findings are consistent with recent experimental observations that suggested that the limit to crystallization pressure is related to the disjoining pressure and not to the thermodynamic limit pressure~\cite{Desarnaud2016,Zheng2018}. Our detailed experiments and modelling indicate that the mechanism responsible for the arrest of crystal growth, is connected to contact healing processes, which create strong but non-covalent adhesion between surfaces confining nanoscale films containing 2-3 layers of water molecules. The new mechanism is strongly dependent on the nature of the surfaces, the roughness and the fluid composition.  Understanding this mechanism will allow prediction of the limit between damage and adhesion by crystallization in many systems in Earth and materials sciences.

%
%
%
%

\begin{acknowledgement}
This project has received funding from the European Union’s Horizon 2020 research and innovation program under the Marie Sklodowska-Curie Grant Agreement No. 642976 (ITN NanoHeal) and from the Norwegian Research Council Grant No. 222386. RMEM acknowledges support of National Science Foundation under the grant Nos. CMMI-1435920 and EAR 18-56525.
\end{acknowledgement}


\begin{suppinfo}
\section{Force applied to crystal\label{sec:force}}
\begin{figure}[ht]
\includegraphics[width=6cm]{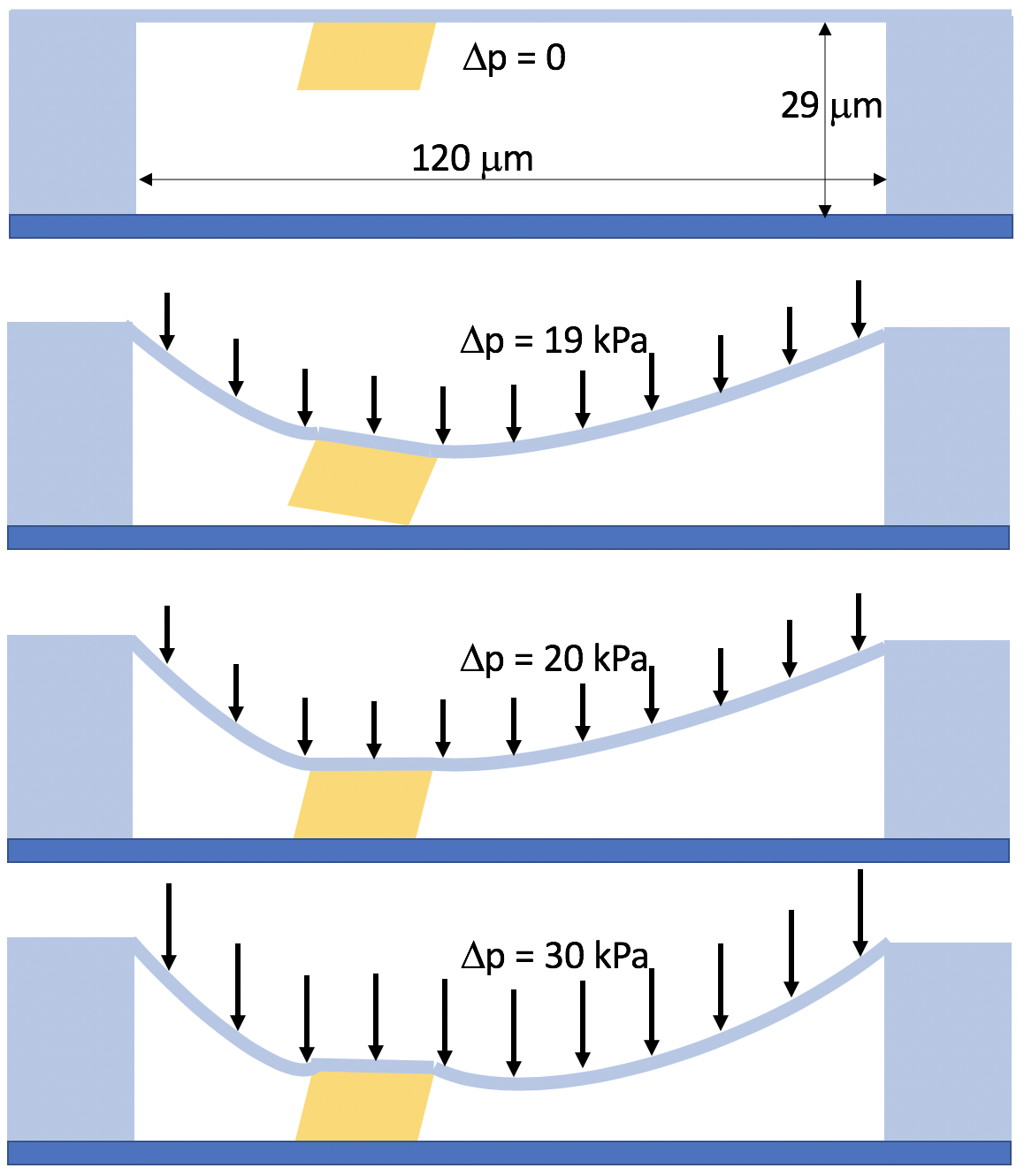}
\includegraphics[width=6cm]{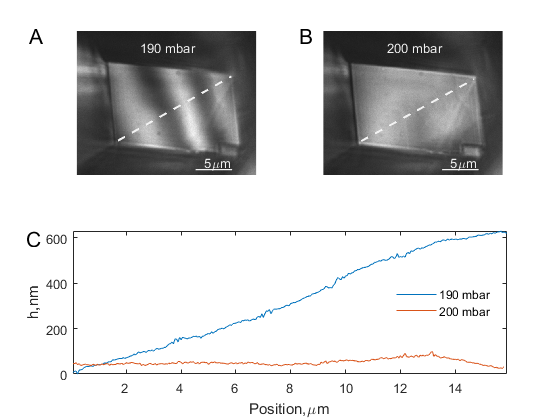}
\caption{Deformation of membrane and force application on crystal. Top: Initially the crystal is suspended underneath the PDMS membrane. The pressure is increased to 19~kPa where one corner of the crystal touches the glass. Further pressure increase to 20~kPa brings the whole crystal surface in contact with the glass. Final pressure increase to 30~kPa does not move crystal, only increases the force between crystal and glass. Bottom part: A and B shows RICM images of crystal brought into contact with glass. C: Distance profile between glass and crystal along whate dashed lines in A and B.}
\label{fig:membrane}
\end{figure}

When the calcite is located and growing on the PDMS membrane in the lower channel, we start to increase the control pressure $P_2$ slowly until a corner of the calcite reaches the cover glass at $P_2=19$~kPa as shown in Figure~\ref{fig:membrane}. The calcite is at this pressure tilted along the white line with a maximum distance of 560nm. In order to bring the calcite surface parallel to the cover glass surface, P$_2$ is increased to 20~kPa as shown in Figure~\ref{fig:membrane}. The average distance along the white line is then 57~nm and the average distance $h$ of the whole surface is 30~nm. As documented previously~\cite{Li2017b} the disjoining pressure $p$ between calcite and a glass surface at distance $h=30$~nm is $p=20\pm 10$~Pa. The force transmitted from the membrane to the crystal is thus the pressure $p$ times the area of the crystal: $F\approx$ 20~Pa$\times$370~$\mu$m$^2\approx$ 7~nN.

This indicates that almost all of the applied fluid pressure $P_2$ is used to deform the PDMS membrane enough to achieve full contact between the calcite and the glass. The pressure change $\Delta P_2=1$~kPa caused an average displacement of the crystal of 280~nm, thus the differential pressure loss in deforming the membrane is $dP_2/dh\approx 3$~Pa/nm. Further fluid pressure increase can displace the crystal at most 30~nm over which the pressure loss in deforming the membrane will be no more than about 90~Pa. 

Since the PDMS membrane did not move the crystal upon the pressure change at $t_1$=90~min, the transmitted pressure increased by $\Delta P_t=\Delta P_2=10$~kPa and the force transmitted from the membrane to the crystal is increased by $F\approx$64~$\mu$N. Since the contact area between the calcite and the glass at this time is only 14~$\mu$m$^2$ the contact pressure increases to about 5~MPa.

\begin{figure}[thb]
\includegraphics[width=8cm]{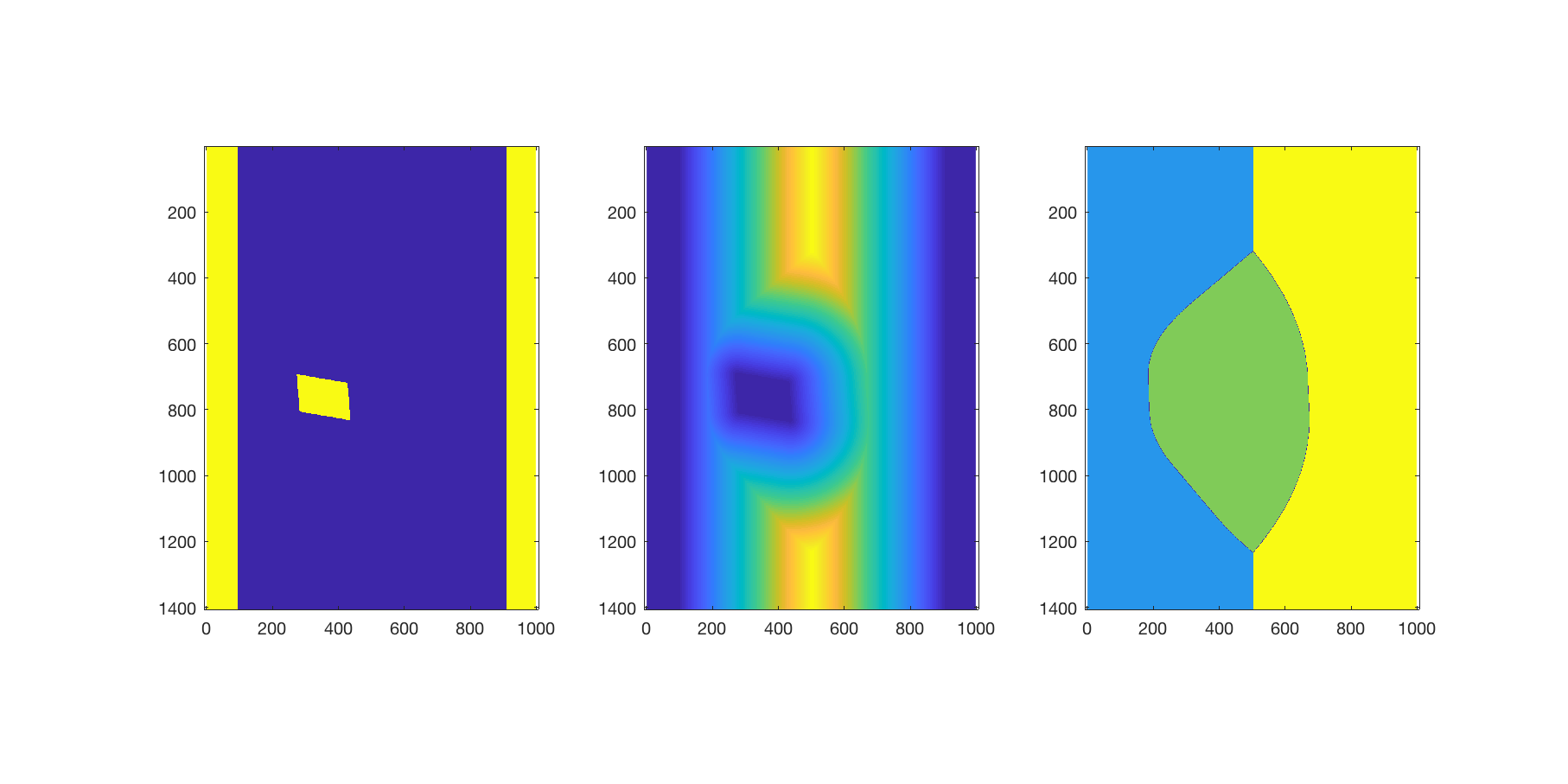}
\caption{Area contributing to force on crystal. Left: Segmented image of crystal in channel. Middle: Euclidean distance transform of the binary image to the left. Right: Watershed showing the area closest to the crystal in green.}
\label{fig:watershed}
\end{figure}
Almost the entire pressure change from 20 to 30~kPa will thus be transferred to the crystal. Multiplying the pressure change by area of the membrane that is closer to the crystal than to the channel walls, $F=\Delta P_2 A_m $=10~kPa$\times$6400~$\mu$m$^2$=64~$\mu$N.

\section{Calcite - silica interactions\label{sec:interactions}}
We combine recent experimental and molecular dynamics data for pressures and diffusion in confined calcite-calcite and calcite-silicate interfaces. We demonstrate how to combine data from experiment and simulation to calculate effective disjoining pressure and diffusion and thereby obtain predictions for deformation rate versus confining pressure.

Recently we have succeeded in measuring surface forces between calcite surfaces and between silica surfaces and calcite surfaces. The last 20 years has also allowed molecular simulation of calcite and silica surfaces with increased refinement and detail. It is not straight forward, however to compare molecular dynamics (MD) simulations, atomic force microscopy (AFM) and surface forces apparatus (SFA) measurements. The individual experiments and simulations are also not easily applied to real problems like colloidal aggregation, material strength, reactivity in confinement, etc. 

The aim of this section is to combine experimental data, molecular simulation and theory and use interaction data from MD and AFM to construct calcite-calcite and calcite-silicate interaction potentials, forces or pressures for flat rough surfaces. In addition we estimate the effect of confinement on diffusion as function of stress and surface roughness. 

\subsection{Interaction energies, forces, pressure, roughness and geometry}
The oscillatory and hydration-steric forces of atomically flat calcite and silica surfaces are as of yet only available from molecular simulation. We can use insights from and models based on measurement of mica-mica forces to extrapolate and add to the DLVO forces. In order to compare this to force measurements on calcite and silica surfaces we need to take the roughness of the surfaces into account.

The free energy of interaction, $u(h)$, between two solid surfaces separated by an electrolyte of thickness $h$ have many contributions. From a theoretical viewpoint one normally assumes that the interaction energy contributions from different parts of the solid surfaces and from different effects in the fluid are all additive. Then one can integrate over the shapes of the surfaces to obtain the force $F(h)=dU(h)/dh$ between the two solids, where $U(h)$ is the free energy  integrated over the surfaces. The force between two spheres of radius $r_1$ and $r_2$ is thus calculated to be~\cite{Israelachvili2011}
\[
F(h)=2\pi\left(\frac{r_1r_2}{r_1+r_2}\right)u(h).
\]
This is called the Derjaguin approximation and is practical to compute forces between two spheres (for example colloidal particles), a sphere and a flat (atomic force microscope (AFM) with a spherical tip on a flat surface) or two cylinders (as in the surface forces apparatus (SFA)). The same assumption of additivity may be used for any surface shape as long as the radii of curvature of the surfaces are much larger than the range of $u(h)$. 

Treating rough surfaces statistically Parsons et al~\cite{Parsons2014} recently proposed performing the corresponding integral over interaction energies using the height probability distributions $p_i(z_i)$ of the two surfaces:
\[
U(h)=\int_{-\infty}^\infty\int_{-\infty}^\infty dz_1dz_2p_1(z_1)p_2(z_2)u(z_1-z_2+h).
\]
This integration is sufficient when the surface roughness, $\sigma_i=(\int_{-\infty}^\infty dz_iz_i^2p_i(z_i))^{1/2}$, is much smaller than the range of $u(h)$. When roughness is larger some points of the surfaces will contact and deform elastically or plastically. Parsons et al~\cite{Parsons2014} proposed to treat the contacting asperities as Hertzian contacts with an effective asperity radius $r_a$. Assuming $p_i(z_i)$ to be Gaussian they can be combined into a single distribution $p(z)$ with variance $\sigma^2=\sqrt{\sigma_1^2+\sigma_2^2}$ and the Hertzian contact contribution, $U_C$, to the interaction between the surfaces is~\cite{Parsons2014}: 
\begin{equation}
U_C(h)=\frac{2E\sigma}{15\pi^{3/2}}\sqrt{\frac{\sigma}{r_a}}\; e^{-\frac{h^2}{2\sigma^2}}f(\frac{h}{\sigma}),
\label{eq:UC}
\end{equation}
where $E$ is an effective Youngs modulus and $f$ is a geometrical function~\cite{Parsons2014}. Both the height probability distributions $p_i(z_i)$ and the typical asperity radius $r_a$ may be estimated from AFM imaging of the surfaces. The total interaction energy, $U_t=U+U_H$, is the sum of the contact and non-contact contributions.

The thermodynamically most useful measure of the surface interactions is the pressure $P(h)=dU(h)/dh/A$, where $A$ is the macroscopic contact area of the surfaces.   

\subsection{Molecular simulation and theory}
We have used the MD data of Brekke-Svaland and Bresme~\cite{Brekke-Svaland2018} that shows oscillatory steric/hydration forces between to flat calcite surfaces in water. We have fit their pressure to $P_{DLVO}+P_{OSH}$ where $P_{OSH}$ is a periodic, exponentially decaying function similar to the suggestion by Israelachvili~\cite{Israelachvili2011} chapter 15.5:
\[
P_{OSH}(h)=P_0\cos(2\pi h/(h_{ml}+\beta h))\exp(-\alpha h/h_{ml}),
\]
with the fitting parameters that are the pressure at contact, $P_0=5\cdot 10^{11}$~Pa, the smallest thickness of a molecular layer of water, $h_{ml}=0.162$~nm, a coefficient of the variation of molecular layer thickness, $\beta=0.03$ and the ratio between the smallest molecular layer thickness and the decay length, $\alpha=1.3$. The pressure contribution from the DLVO theory for flat surfaces, $P_{DLVO}$ was calculated using the parameters published by Diao and Espinosa-Marzal~\cite{Diao2016}.

\subsection{Disjoining pressure model for rough silica on flat calcite}
In order to combine molecular dynamics, theory and experiment we have combined the  model fit in the previous section with experimental parameters and data from Diao and Espinoza-Marzal~\cite{Diao2016}. They reported that the roughness of the silica spheres was approximately $\sigma=$2~nm. To model this roughness we used a truncated normal distribution
\[
n(z)=\begin{cases}\frac{1}{N} e^{-\frac{z^2}{2\sigma^2}}, & |z|<2\sigma\\
=0, & |z|>2\sigma\end{cases},
\]
where the $N=\int_{-2}^2dz e^{-\frac{z^2}{2\sigma^2}}$ normalizes $n$ as a probability distribution. The pressures $P_{R,i}$ for the rough system are then
\[
P_{R,i}(h)=\int_{-2\sigma}^{2\sigma} dz P_i(h-z)n(z).
\]

The elastic contact contribution to the pressure is found by differentiating equation (\ref{eq:UC}):
\[
P_C(h)=\frac{2E\sigma}{15\pi^{3/2}}\sqrt{\frac{\sigma}{r_a}}\; e^{-\frac{h^2}{2\sigma^2}}
\left(\frac{h}{\sigma}f(\frac{h}{\sigma})-f'(\frac{h}{\sigma})\right).
\]

The experimental data of Diao and Espinosa-Marzal~\cite{Diao2016} for 0.51~mM CaCO$_3$ solution that has been translated along the x-axis to coincide with the pressure model. The shift is justified because the AFM data has no intrinsic reference point of $x=0$. In order to obtain a region of the model with a moderate slope corresponding to the data, the roughness and asperity radius were adjusted to $\sigma=3$~nm and $r_a=10$~nm. 

We have also included data from Li et al~\cite{Li2017b} of a crystal lying on a glass surface with roughness of about 0.2~nm. This data agrees perfectly with the roughness corrected pressure model presented here.

\section{Diffusion\label{sec:diffusion}}
\begin{figure}[ht]
\centering
  \includegraphics[width=8cm]{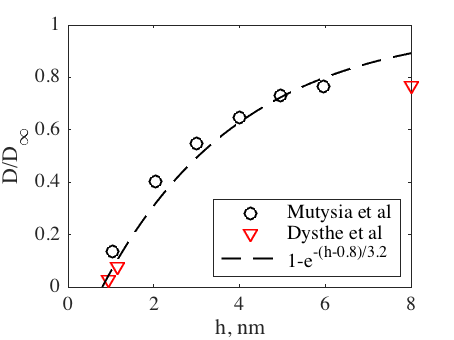}
  \caption{Self diffusion of water in calcite slit pores as function of the water film thickness, $h$.}
  \label{fig:DoverDinfty}
\end{figure}
Self diffusion of water in a calcite slit pore has been calculated by Dysthe et al~\cite{Dysthe2002} and Mutysia et al~\cite{Mutisya2017}. The interaction potential model used by Mutysia et al~\cite{Mutisya2017} is more evolved and adapted to a range of properties of calcite-water interfaces. A rough empirical fit to their data for water, $D_w$, yields $D_w/D_{w,\infty} =1-\exp(-(h-h_0)/\lambda)$ with $h_0$=0.8~nm and $\lambda$=3.2~nm and $D_{w,\infty}$ is the bulk diffusion coefficient.  Figure~\ref{fig:DoverDinfty} shows that the diffusion reduced coefficients $D_w/D_{w,\infty}$ calculated by Dysthe et al~\cite{Dysthe2002} with the simplified interaction potential at strongly varying temperatures agree well with the data of Mutysia et al~\cite{Mutisya2017}. 

We are interested in the reactivity of a nanoconfined calcite interface and how it is limited by mass transport. The quantities of interest is then the interdiffusion coefficient of CaCO$_3$ that depends on the self diffusion coefficients of the Ca$^{2+}$ and CO$_3^{2-}$ ions. If we assume that the Stokes-Einstein relation $D=kT/6\pi\eta r$, where $\eta$ is the viscosity and $r$ is the particle radius, is valid for both water and ions. That is, the water self diffusion measurements reflect the change in viscosity of the fluid layer and that the mobility of all ions are inversely proportional to this viscosity. Then the calcium carbonate diffusion coefficient, $D_c$ should also be $D_c/D_{c,\infty} =1-\exp(-(h-0.8)/3.2)$.

Diffusion in silica slit pores has been studied by Collin et al~\cite{Collin2018} and they found that similarly to calcite water diffusion is reduced by an order of magnitude for slit pores of 1~nm corresponding to 3 water layers.

One important aspect of a rough surface meeting a flat surface is that it leaves room for diffusion. At a high pressure like 1~GPa diffusion is very slow between two smooth, flat surfaces. For the rough silica - flat calcite contact at similar pressures only a small part of the surface has a distance $h<0.8$~nm where diffusion is zero. Unless the reactive surface changes shape to conform with the rough surface, diffusion will continue and so will crystal growth.

\section{Local contact formation modelled by Kinetic Monte Carlo\label{sec:KMC}}
\begin{figure*}[ht]
\centering
  \includegraphics[width=12cm]{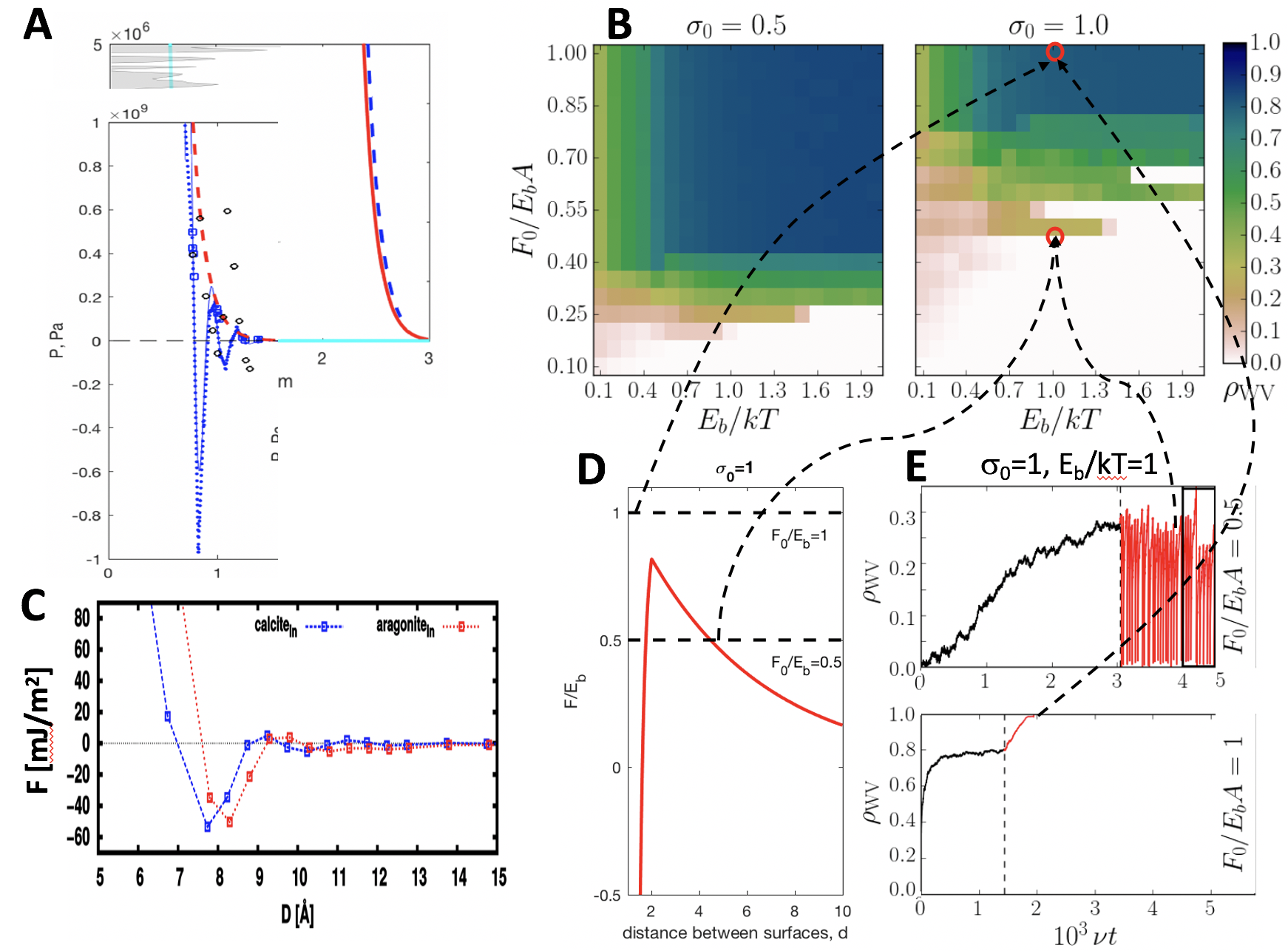}
  \caption{Growing contacts under global pressure and local attraction. {\bf A:} Global repulsion between the calcite and the rough glass surface and the local oscillatory pressures (note different vertical scales). Image from this study. {\bf D:} Repulsive and attractive pressure used in KMC model. {\bf B and E:} Results of KMC simulations showing complete contact developing ($\rho_{WV}=1$) when the load is larger than disjoining pressure peak and partial coverage and oscillatory behaviour when the load is smaller than the disjoining pressure and the equilibrium intersurface distance is 3 crystal layers. Images from~\cite{Hogberget2020a}. {\bf C:} Free energy of interaction between two flat calcite surfaces. The depth of the minimum due to ordering of 2 water layers is $F/kT=2$. Figure reproduced from~\cite{Brekke-Svaland2018}.
  }
  \label{fig:KMC}
\end{figure*}
A recent study of contact formation using Kinetic Monte Carlo (KMC)\cite{Hogberget2020a} gives some insight into how local dissolution and growth form adhesive contacts between a reactive and a non-reactive surface. The model uses an exponential surface-surface repulsive energy $G_r=E_b\sigma_0 e^{-(h-h_0)/\lambda_D}$, where $E_b$ is the depth of the energy minimum binding the two solids, $\sigma_0$ is the strength of the repulsive interaction, $\lambda_D$ is the Debye length and $h_0$ is the position of minimum of the attractive potential $G_a=-E_b(3h+64/h^6-7)/60, h\leq h_0$. The KMC simulation of an atomically flat non-reactive confining surface and a reactive surface with a loading pressure $F_0/A$ can be compared to the experimental system in the following way:

There is a global loading force and a global disjoining force that balance as shown in the main text and in Figure \ref{fig:KMC}{\bf A}. In the experiments there are local asperities of the glass surface where fluctuations in the local height of the calcite may bring the two as close as 2 or 3 layers of water where there is attractive energy between the two surfaces (see Figure \ref{fig:KMC}{\bf C}). The depth of this energy minimum is mainly controlled by the ordering of the water layers between the surfaces. This is sensitive to the molecular structure of the surfaces and the relative positioning of the crystal lattices and for calcite-calcite $F$=-20 to -60kJ/m$^2$=-3.2 to -8.1$\cdot 10^{-21}$J per lattice site, thus $-F/kT=0.8-2$. If the crystals are ordered such that the water layer ordering is completely broken the main minimum will be direct solid-solid contact (no water left in the contact).

This global repulsion, and local attraction resembles the KMC model as shown in Figure \ref{fig:KMC}{\bf D}. At lower loads the system can end up in an oscillating regime where contacts are formed and broken like we observe in the initial low load part of the experiments. As the load is increased there is a transition to full contact developing between the two surfaces. In our system this contact state may correspond to a local surface-surface distance of either 2-3 molecular water layers or to zero distance, that is solid-solid contact.

\section{SFA experiments}
\begin{figure*}[ht]
\centering
  \includegraphics[width=12cm]{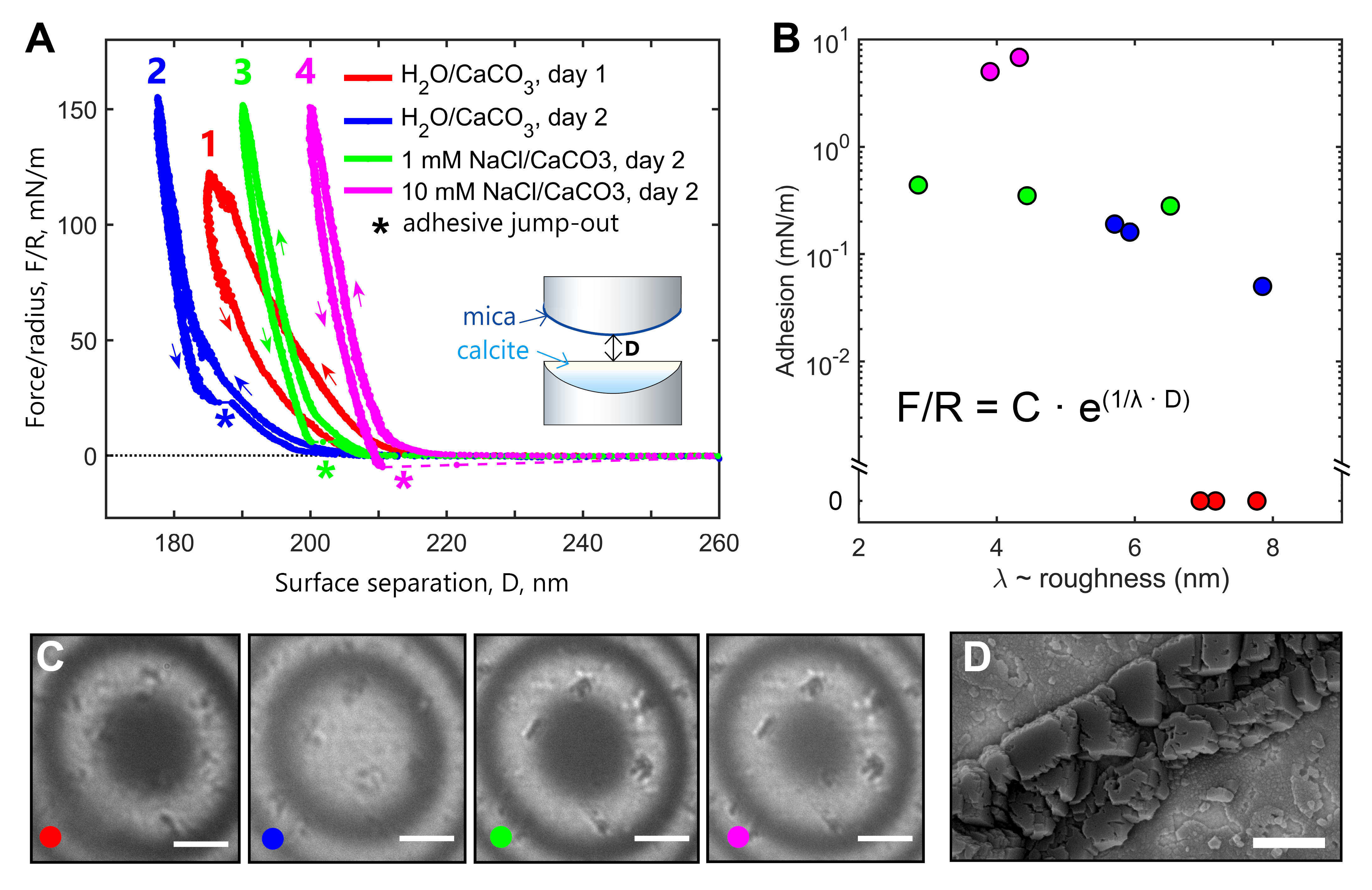}
  \caption{SFA measurements of forces between reactive calcite and inert mica surfaces indicating a progressive increase in the local contact area. {\bf A:} Representative force-separation distance curves measured between rough calcite and smooth mica surfaces in water and NaCl solutions saturated with respect to calcite. The numbers mark the order of measurements. Upon recrystallization, the calcite surfaces progressively grow as shown by the increase in the minimum surface separation (apart from the initial dissolution between solutions 1 and 2). Despite the growth, the local contact area becomes smoother as indicated by the increasing adhesion and decreasing exponential decay length of the force curves measured on approach (see panel B). {\bf B:} Exponential decay length ($\lambda$) of the SFA force-distance curves measured on approach ($\lambda$ is proportional to the local contact roughness) as a function of adhesion between mica and calcite surfaces (for color-coding see the legend in A). {\bf C:} In-situ optical view from the SFA on the contact topography showing the growth of larger calcite asperities in the contact region between two surfaces (outlined by visible Newton rings). The scale bars are 50 $\mu$m. {\bf D:} Scanning electron microscopy (SEM) image of one of the larger calcite asperities, showing flat regions on the recrystallized calcite surface. The scale bar is 1 $\mu$m.
  }
  \label{fig:SFA-mc}
\end{figure*}

We observed an analogous process, in which the reactive calcite surface grows locally in confinement to increase the adhesive surface area with the opposing solid surface, in the Surface Forces Apparatus (SFA). We used one rough and reactive calcite surface against a smooth and inert muscovite mica surface (see Figure \ref{fig:SFA-mc}). The forces measured in the SFA allowed monitoring of the changing contact roughness and surface adhesion during the reactive surface growth. 

The preparation of surfaces and details of the SFA experiments have been previously described in Dziadkowiec et al \cite{Dziadkowiec2018}. We used polycrystalline calcite surfaces grown by atomic layer deposition (ALD). Despite using calcite-saturated solutions at all times, we observed recrystallization of calcite in contact with aqueous solutions, which was driven by the disequilibrium morphology of the ALD-deposited calcite crystals \cite{Dziadkowiec2019}. The SFA measurements were performed in micron-scale confinement: the distance between the surfaces was at all times $<$ 1 $\mu$m in the most confined region (surfaces were placed in a crossed-cylindrical geometry, yielding a spherical confined contact area with a diameter of $\sim$150 $\mu$m).

Figure \ref{fig:SFA-mc}{\bf A} shows a sequence of force-distance curves measured between mica and calcite surfaces in the same contact region. After initial dissolution (solutions 1 to 2), we observed a progressive growth of calcite surfaces (reflected by the increasing minimum surface separation) associated with the decrease in the repulsive force components (indicated by the decreasing exponential decay length of the force curves; Figure \ref{fig:SFA-mc}{\bf B}) and the increasing adhesion. 

The forces measured in the SFA were dominated by the repulsive roughness contribution \cite{Parsons2014}. Thus, the major changes in the repulsive decay length ($\lambda$) of the force curves can be almost entirely attributed to the changing 'local' roughness of the calcite surface asperities \cite{Dziadkowiec2019} and not to the slight variations in solution chemistry. According to Benz et al \cite{benz2006deformation}, the local contact roughness ($\sigma$) is proportional to the exponential decay length ($\lambda$) of the force (F)-distance (D) curves measured on approach according to: $F/R=C*e^{(-D/\lambda)}$, where $\lambda = 2*\sigma$, and C is a fitting constant. Although, based on the optical in-situ images of the contact region (Figure \ref{fig:SFA-mc}{\bf C}), we can directly observe that the overall roughness of the calcite surfaces was increasing (see dark large calcite asperities appearing with time), the local roughness of the calcite asperities in contact with mica decreased. This smoothing of the contacts led to the increase in an adhesive contact area, as evidenced by the decreasing $\lambda$ and the increasing adhesion. The smooth faces on the recrystallized calcite asperities could be evidenced ex-situ with Scanning Electron Microscopy (SEM; Figure \ref{fig:SFA-mc}{\bf D}). As such, calcite growing in the confined SFA geometry did not exert pressure on the opposing mica surface (which would be evidenced by the increase in repulsive force contribution during the force measurements). Instead, the calcite crystals grew locally to maximize the adhesive contact area with mica. 

\end{suppinfo}

\bibliography{limitsFoX}

\providecommand{\latin}[1]{#1}
\makeatletter
\providecommand{\doi}
  {\begingroup\let\do\@makeother\dospecials
  \catcode`\{=1 \catcode`\}=2 \doi@aux}
\providecommand{\doi@aux}[1]{\endgroup\texttt{#1}}
\makeatother
\providecommand*\mcitethebibliography{\thebibliography}
\csname @ifundefined\endcsname{endmcitethebibliography}
  {\let\endmcitethebibliography\endthebibliography}{}
\begin{mcitethebibliography}{49}
\providecommand*\natexlab[1]{#1}
\providecommand*\mciteSetBstSublistMode[1]{}
\providecommand*\mciteSetBstMaxWidthForm[2]{}
\providecommand*\mciteBstWouldAddEndPuncttrue
  {\def\EndOfBibitem{\unskip.}}
\providecommand*\mciteBstWouldAddEndPunctfalse
  {\let\EndOfBibitem\relax}
\providecommand*\mciteSetBstMidEndSepPunct[3]{}
\providecommand*\mciteSetBstSublistLabelBeginEnd[3]{}
\providecommand*\EndOfBibitem{}
\mciteSetBstSublistMode{f}
\mciteSetBstMaxWidthForm{subitem}{(\alph{mcitesubitemcount})}
\mciteSetBstSublistLabelBeginEnd
  {\mcitemaxwidthsubitemform\space}
  {\relax}
  {\relax}

\bibitem[Espinosa-Marzal and Scherer(2010)Espinosa-Marzal, and
  Scherer]{Espinosa-Marzal2010}
Espinosa-Marzal,~R.~M.; Scherer,~G.~W. {Advances in understanding damage by
  salt crystallization.} \emph{Accounts of chemical research} \textbf{2010},
  \emph{43}, 897--905\relax
\mciteBstWouldAddEndPuncttrue
\mciteSetBstMidEndSepPunct{\mcitedefaultmidpunct}
{\mcitedefaultendpunct}{\mcitedefaultseppunct}\relax
\EndOfBibitem
\bibitem[Flatt \latin{et~al.}(2014)Flatt, Caruso, Sanchez, and
  Scherer]{Flatt2014a}
Flatt,~R.~J.; Caruso,~F.; Sanchez,~A. M.~A.; Scherer,~G.~W. {Chemo-mechanics of
  salt damage in stone.} \emph{Nature communications} \textbf{2014}, \emph{5},
  4823\relax
\mciteBstWouldAddEndPuncttrue
\mciteSetBstMidEndSepPunct{\mcitedefaultmidpunct}
{\mcitedefaultendpunct}{\mcitedefaultseppunct}\relax
\EndOfBibitem
\bibitem[Gratier \latin{et~al.}(2012)Gratier, Frery, Deschamps, R{\o}yne,
  Renard, Dysthe, Ellouz-Zimmerman, and Hamelin]{Gratier2012a}
Gratier,~J.~P.; Frery,~E.; Deschamps,~P.; R{\o}yne,~A.; Renard,~F.; Dysthe,~D.;
  Ellouz-Zimmerman,~N.; Hamelin,~B. {How travertine veins grow from top to
  bottom and lift the rocks above them: The effect of crystallization force}.
  \emph{Geology} \textbf{2012}, \emph{40}, 1015--1018\relax
\mciteBstWouldAddEndPuncttrue
\mciteSetBstMidEndSepPunct{\mcitedefaultmidpunct}
{\mcitedefaultendpunct}{\mcitedefaultseppunct}\relax
\EndOfBibitem
\bibitem[Taber(1928)]{Taber1928}
Taber,~S. {The growth of crystals under external pressure}. \emph{American
  Journal of Science} \textbf{1928}, \emph{41}, 532\relax
\mciteBstWouldAddEndPuncttrue
\mciteSetBstMidEndSepPunct{\mcitedefaultmidpunct}
{\mcitedefaultendpunct}{\mcitedefaultseppunct}\relax
\EndOfBibitem
\bibitem[Wiltschko and Morse(2001)Wiltschko, and Morse]{Wiltschko2001}
Wiltschko,~D.~V.; Morse,~J.~W. {Crystallization pressure versus "crack seal" as
  the mechanism for banded veins}. \emph{Geology} \textbf{2001}, \emph{29},
  79--82\relax
\mciteBstWouldAddEndPuncttrue
\mciteSetBstMidEndSepPunct{\mcitedefaultmidpunct}
{\mcitedefaultendpunct}{\mcitedefaultseppunct}\relax
\EndOfBibitem
\bibitem[R{\o}yne \latin{et~al.}(2008)R{\o}yne, Jamtveit, Mathiesen, and
  Malthe-S{\o}renssen]{Royne2008}
R{\o}yne,~A.; Jamtveit,~B.; Mathiesen,~J.; Malthe-S{\o}renssen,~A. {Controls on
  rock weathering rates by reaction-induced hierarchical fracturing}.
  \emph{Earth and Planetary Science Letters} \textbf{2008}, \emph{275},
  364--369\relax
\mciteBstWouldAddEndPuncttrue
\mciteSetBstMidEndSepPunct{\mcitedefaultmidpunct}
{\mcitedefaultendpunct}{\mcitedefaultseppunct}\relax
\EndOfBibitem
\bibitem[Kelemen and Hirth(2012)Kelemen, and Hirth]{Kelemen2012}
Kelemen,~P.~B.; Hirth,~G. {Reaction-driven cracking during retrograde
  metamorphism: Olivine hydration and carbonation}. \emph{Earth and Planetary
  Science Letters} \textbf{2012}, \emph{345-348}, 81--89\relax
\mciteBstWouldAddEndPuncttrue
\mciteSetBstMidEndSepPunct{\mcitedefaultmidpunct}
{\mcitedefaultendpunct}{\mcitedefaultseppunct}\relax
\EndOfBibitem
\bibitem[Rodriguez-Navarro \latin{et~al.}(2000)Rodriguez-Navarro, Doehne, and
  Sebastian]{Rodriguez-Navarro2000}
Rodriguez-Navarro,~C.; Doehne,~E.; Sebastian,~E. {Influencing crystallization
  damage in porous materials through the use of surfactants: experimental
  results using sodium dodecyl sulfate and cetyldimethylbenzylammonium
  chloride}. \emph{Langmuir} \textbf{2000}, \emph{16}, 947--954\relax
\mciteBstWouldAddEndPuncttrue
\mciteSetBstMidEndSepPunct{\mcitedefaultmidpunct}
{\mcitedefaultendpunct}{\mcitedefaultseppunct}\relax
\EndOfBibitem
\bibitem[Desarnaud \latin{et~al.}(2016)Desarnaud, Bonn, and
  Shahidzadeh]{Desarnaud2016}
Desarnaud,~J.; Bonn,~D.; Shahidzadeh,~N. {The Pressure induced by salt
  crystallization in confinement}. \emph{Nature Publishing Group}
  \textbf{2016}, \emph{6}, 23--26\relax
\mciteBstWouldAddEndPuncttrue
\mciteSetBstMidEndSepPunct{\mcitedefaultmidpunct}
{\mcitedefaultendpunct}{\mcitedefaultseppunct}\relax
\EndOfBibitem
\bibitem[Jia \latin{et~al.}(2019)Jia, Liang, He, Zhao, and Simon]{Jia2019}
Jia,~M.; Liang,~J.; He,~L.; Zhao,~X.; Simon,~S. {Hydrophobic and hydrophilic
  SiO2-based hybrids in the protection of sandstone for anti-salt damage}.
  \emph{Journal of Cultural Heritage} \textbf{2019}, \emph{40}, 80--91\relax
\mciteBstWouldAddEndPuncttrue
\mciteSetBstMidEndSepPunct{\mcitedefaultmidpunct}
{\mcitedefaultendpunct}{\mcitedefaultseppunct}\relax
\EndOfBibitem
\bibitem[Zheng \latin{et~al.}(2018)Zheng, Cordonnier, Zhu, Renard, and
  Jamtveit]{Zheng2018}
Zheng,~X.; Cordonnier,~B.; Zhu,~W.; Renard,~F.; Jamtveit,~B. {Effects of
  Confinement on Reaction-Induced Fracturing During Hydration of Periclase}.
  \emph{Geochemistry, Geophysics, Geosystems} \textbf{2018}, \emph{19},
  2661--2672\relax
\mciteBstWouldAddEndPuncttrue
\mciteSetBstMidEndSepPunct{\mcitedefaultmidpunct}
{\mcitedefaultendpunct}{\mcitedefaultseppunct}\relax
\EndOfBibitem
\bibitem[Becker and Day(1905)Becker, and Day]{Becker1905}
Becker,~G.~F.; Day,~A.~L. {The linerar force of growing crystals}. \emph{Proc.
  Washingt. Acad. Sci.} \textbf{1905}, \emph{7}, 283--288\relax
\mciteBstWouldAddEndPuncttrue
\mciteSetBstMidEndSepPunct{\mcitedefaultmidpunct}
{\mcitedefaultendpunct}{\mcitedefaultseppunct}\relax
\EndOfBibitem
\bibitem[Becker and Day(1916)Becker, and Day]{Becker1916}
Becker,~G.~F.; Day,~A.~L. {Note on the Linear Force of Growing Crystals}.
  \emph{Source: The Journal of Geology} \textbf{1916}, 313--333\relax
\mciteBstWouldAddEndPuncttrue
\mciteSetBstMidEndSepPunct{\mcitedefaultmidpunct}
{\mcitedefaultendpunct}{\mcitedefaultseppunct}\relax
\EndOfBibitem
\bibitem[Flatt \latin{et~al.}(2007)Flatt, Steiger, and Scherer]{Flatt2007}
Flatt,~R.~J.; Steiger,~M.; Scherer,~G.~W. {A commented translation of the paper
  by C.W. Correns and W. Steinborn on crystallization pressure}.
  \emph{Environmental Geology} \textbf{2007}, \emph{52}, 221--237\relax
\mciteBstWouldAddEndPuncttrue
\mciteSetBstMidEndSepPunct{\mcitedefaultmidpunct}
{\mcitedefaultendpunct}{\mcitedefaultseppunct}\relax
\EndOfBibitem
\bibitem[Sekine \latin{et~al.}(2011)Sekine, Okamoto, and Hayashi]{Sekine2011}
Sekine,~K.; Okamoto,~a.; Hayashi,~K. {In situ observation of the
  crystallization pressure induced by halite crystal growth in a microfluidic
  channel}. \emph{American Mineralogist} \textbf{2011}, \emph{96},
  1012--1019\relax
\mciteBstWouldAddEndPuncttrue
\mciteSetBstMidEndSepPunct{\mcitedefaultmidpunct}
{\mcitedefaultendpunct}{\mcitedefaultseppunct}\relax
\EndOfBibitem
\bibitem[R{\o}yne and Dysthe(2012)R{\o}yne, and Dysthe]{Royne2012b}
R{\o}yne,~A.; Dysthe,~D.~K. {Rim formation on crystal faces growing in
  confinement}. \emph{Journal of Crystal Growth} \textbf{2012}, \emph{346},
  89--100\relax
\mciteBstWouldAddEndPuncttrue
\mciteSetBstMidEndSepPunct{\mcitedefaultmidpunct}
{\mcitedefaultendpunct}{\mcitedefaultseppunct}\relax
\EndOfBibitem
\bibitem[Naillon \latin{et~al.}(2018)Naillon, Joseph, and Prat]{Naillon2018}
Naillon,~A.; Joseph,~P.; Prat,~M. {Ion Transport and Precipitation Kinetics as
  Key Aspects of Stress Generation on Pore Walls Induced by Salt
  Crystallization}. \emph{Physical Review Letters} \textbf{2018}, \emph{120},
  034502\relax
\mciteBstWouldAddEndPuncttrue
\mciteSetBstMidEndSepPunct{\mcitedefaultmidpunct}
{\mcitedefaultendpunct}{\mcitedefaultseppunct}\relax
\EndOfBibitem
\bibitem[Kohler \latin{et~al.}(2018)Kohler, Gagliardi, Pierre-Louis, and
  Dysthe]{Kohler2018}
Kohler,~F.; Gagliardi,~L.; Pierre-Louis,~O.; Dysthe,~D.~K. {Cavity Formation in
  Confined Growing Crystals}. \emph{Physical Review Letters} \textbf{2018},
  \emph{121}, 96101\relax
\mciteBstWouldAddEndPuncttrue
\mciteSetBstMidEndSepPunct{\mcitedefaultmidpunct}
{\mcitedefaultendpunct}{\mcitedefaultseppunct}\relax
\EndOfBibitem
\bibitem[Li \latin{et~al.}(2018)Li, Sanchez, Kohler, R{\o}yne, and
  Dysthe]{Li2018}
Li,~L.; Sanchez,~J.~R.; Kohler,~F.; R{\o}yne,~A.; Dysthe,~D.~K. {Microfluidic
  Control of Nucleation and Growth of CaCO3}. \emph{Crystal Growth and Design}
  \textbf{2018}, \emph{18}, 4528--4535\relax
\mciteBstWouldAddEndPuncttrue
\mciteSetBstMidEndSepPunct{\mcitedefaultmidpunct}
{\mcitedefaultendpunct}{\mcitedefaultseppunct}\relax
\EndOfBibitem
\bibitem[Weyl(1959)]{Weyl1959}
Weyl,~P.~K. {Pressure solution and the force of crystallization: a
  phenomenological theory}. \emph{Journal of Geophysical Research}
  \textbf{1959}, \emph{64}, 2001--2025\relax
\mciteBstWouldAddEndPuncttrue
\mciteSetBstMidEndSepPunct{\mcitedefaultmidpunct}
{\mcitedefaultendpunct}{\mcitedefaultseppunct}\relax
\EndOfBibitem
\bibitem[Li \latin{et~al.}(2017)Li, Kohler, R{\o}yne, and Dysthe]{Li2017b}
Li,~L.; Kohler,~F.; R{\o}yne,~A.; Dysthe,~D. {Growth of Calcite in
  Confinement}. \emph{Crystals} \textbf{2017}, \emph{7}, 361\relax
\mciteBstWouldAddEndPuncttrue
\mciteSetBstMidEndSepPunct{\mcitedefaultmidpunct}
{\mcitedefaultendpunct}{\mcitedefaultseppunct}\relax
\EndOfBibitem
\bibitem[Teng \latin{et~al.}(2000)Teng, Dove, and {De Yoreo}]{Teng2000}
Teng,~H.~H.; Dove,~P.~M.; {De Yoreo},~J.~J. {Kinetics of calcite growth:
  Surface processes and relationships to macroscopic rate laws}.
  \emph{Geochimica et Cosmochimica Acta} \textbf{2000}, \emph{64},
  2255--2266\relax
\mciteBstWouldAddEndPuncttrue
\mciteSetBstMidEndSepPunct{\mcitedefaultmidpunct}
{\mcitedefaultendpunct}{\mcitedefaultseppunct}\relax
\EndOfBibitem
\bibitem[Charlton and Parkhurst(2011)Charlton, and Parkhurst]{Charlton2011}
Charlton,~S.~R.; Parkhurst,~D.~L. {Modules based on the geochemical model
  PHREEQC for use in scripting and programming languages}. \emph{Computers and
  Geosciences} \textbf{2011}, \emph{37}, 1653--1663\relax
\mciteBstWouldAddEndPuncttrue
\mciteSetBstMidEndSepPunct{\mcitedefaultmidpunct}
{\mcitedefaultendpunct}{\mcitedefaultseppunct}\relax
\EndOfBibitem
\bibitem[Dysthe(2014)]{Dysthe2014}
Dysthe,~D.~K. In \emph{Transport and Reactivity of Solutions in Confined
  Hydrosystems}; Mercury,~L., Tas,~N., Zilberbrand,~M., Eds.; Springer Science
  + Business Media: Dordrecht, 2014; Chapter 17, p 199\relax
\mciteBstWouldAddEndPuncttrue
\mciteSetBstMidEndSepPunct{\mcitedefaultmidpunct}
{\mcitedefaultendpunct}{\mcitedefaultseppunct}\relax
\EndOfBibitem
\bibitem[Supplemetary-Information()]{SupInf}
Supplemetary-Information, {See the Supplementary Information}. \emph{2020}
  \relax
\mciteBstWouldAddEndPunctfalse
\mciteSetBstMidEndSepPunct{\mcitedefaultmidpunct}
{}{\mcitedefaultseppunct}\relax
\EndOfBibitem
\bibitem[Brekke-Svaland and Bresme(2018)Brekke-Svaland, and
  Bresme]{Brekke-Svaland2018}
Brekke-Svaland,~G.; Bresme,~F. {Interactions between Hydrated Calcium Carbonate
  Surfaces at Nanoconfinement Conditions}. \emph{Journal of Physical Chemistry
  C} \textbf{2018}, \emph{122}, 7321--7330\relax
\mciteBstWouldAddEndPuncttrue
\mciteSetBstMidEndSepPunct{\mcitedefaultmidpunct}
{\mcitedefaultendpunct}{\mcitedefaultseppunct}\relax
\EndOfBibitem
\bibitem[Dysthe \latin{et~al.}(2002)Dysthe, Renard, Porcheron, and
  Rousseau]{Dysthe2002}
Dysthe,~D.~K.; Renard,~F.; Porcheron,~F.; Rousseau,~B. {Fluid in mineral
  interfaces—molecular simulations of structure and diffusion}.
  \emph{Geophysical Research Letters} \textbf{2002}, \emph{29}, 13--14\relax
\mciteBstWouldAddEndPuncttrue
\mciteSetBstMidEndSepPunct{\mcitedefaultmidpunct}
{\mcitedefaultendpunct}{\mcitedefaultseppunct}\relax
\EndOfBibitem
\bibitem[Diao and Espinosa-Marzal(2016)Diao, and Espinosa-Marzal]{Diao2016}
Diao,~Y.; Espinosa-Marzal,~R.~M. {Molecular insight into the nanoconfined
  calcite–solution interface}. \emph{Proceedings of the National Academy of
  Sciences} \textbf{2016}, \emph{113}, 12047--12052\relax
\mciteBstWouldAddEndPuncttrue
\mciteSetBstMidEndSepPunct{\mcitedefaultmidpunct}
{\mcitedefaultendpunct}{\mcitedefaultseppunct}\relax
\EndOfBibitem
\bibitem[Israelachvili(2011)]{Israelachvili2011}
Israelachvili,~J.~N. \emph{{Intermolecular and surface forces}}; Academic
  Press, 2011; p 674\relax
\mciteBstWouldAddEndPuncttrue
\mciteSetBstMidEndSepPunct{\mcitedefaultmidpunct}
{\mcitedefaultendpunct}{\mcitedefaultseppunct}\relax
\EndOfBibitem
\bibitem[Parsons \latin{et~al.}(2014)Parsons, Walsh, and Craig]{Parsons2014}
Parsons,~D.~F.; Walsh,~R.~B.; Craig,~V.~S. {Surface forces: Surface roughness
  in theory and experiment}. \emph{Journal of Chemical Physics} \textbf{2014},
  \emph{140}\relax
\mciteBstWouldAddEndPuncttrue
\mciteSetBstMidEndSepPunct{\mcitedefaultmidpunct}
{\mcitedefaultendpunct}{\mcitedefaultseppunct}\relax
\EndOfBibitem
\bibitem[Freitas and Reed(2020)Freitas, and Reed]{Freitas2020}
Freitas,~R.; Reed,~E.~J. {Uncovering the effects of interface-induced ordering
  of liquid on crystal growth using machine learning}. \emph{Nature
  Communications} \textbf{2020}, \emph{11}, 1--10\relax
\mciteBstWouldAddEndPuncttrue
\mciteSetBstMidEndSepPunct{\mcitedefaultmidpunct}
{\mcitedefaultendpunct}{\mcitedefaultseppunct}\relax
\EndOfBibitem
\bibitem[C{\'{a}}mara and Bresme(2004)C{\'{a}}mara, and Bresme]{Camara2004}
C{\'{a}}mara,~L.~G.; Bresme,~F. {Liquids confined in wedge shaped pores:
  Nonuniform pressure induced by pore geometry}. \emph{Journal of Chemical
  Physics} \textbf{2004}, \emph{120}, 11355--11358\relax
\mciteBstWouldAddEndPuncttrue
\mciteSetBstMidEndSepPunct{\mcitedefaultmidpunct}
{\mcitedefaultendpunct}{\mcitedefaultseppunct}\relax
\EndOfBibitem
\bibitem[H{\o}gberget \latin{et~al.}(2020)H{\o}gberget, R{\o}yne, Dysthe, and
  Jettestuen]{Hogberget2020a}
H{\o}gberget,~J.; R{\o}yne,~A.; Dysthe,~D.~K.; Jettestuen,~E. {Microscopic
  modeling of contact formation between confined surfaces in solution}.
  \emph{arXiv:cond-mat} \textbf{2020}, \emph{2006.02129}, 1--16\relax
\mciteBstWouldAddEndPuncttrue
\mciteSetBstMidEndSepPunct{\mcitedefaultmidpunct}
{\mcitedefaultendpunct}{\mcitedefaultseppunct}\relax
\EndOfBibitem
\bibitem[R{\o}yne \latin{et~al.}(2015)R{\o}yne, Dalby, and
  Hassenkam]{Royne2015}
R{\o}yne,~A.; Dalby,~K.~N.; Hassenkam,~T. {Repulsive hydration forces between
  calcite surfaces and their effect on the brittle strength of calcite-bearing
  rocks}. \emph{Geophysical Research Letters} \textbf{2015}, \emph{42},
  4786--4794\relax
\mciteBstWouldAddEndPuncttrue
\mciteSetBstMidEndSepPunct{\mcitedefaultmidpunct}
{\mcitedefaultendpunct}{\mcitedefaultseppunct}\relax
\EndOfBibitem
\bibitem[Dziadkowiec \latin{et~al.}(2018)Dziadkowiec, Javadi, Bratvold, Nilsen,
  and R{\o}yne]{Dziadkowiec2018}
Dziadkowiec,~J.; Javadi,~S.; Bratvold,~J.; Nilsen,~O.; R{\o}yne,~A. {Surface
  Forces Apparatus Measurements of Interactions between Rough and Reactive
  Calcite Surfaces}. \emph{Langmuir} \textbf{2018}, \emph{34}\relax
\mciteBstWouldAddEndPuncttrue
\mciteSetBstMidEndSepPunct{\mcitedefaultmidpunct}
{\mcitedefaultendpunct}{\mcitedefaultseppunct}\relax
\EndOfBibitem
\bibitem[Javadi and R{\o}yne(2018)Javadi, and R{\o}yne]{Javadi2018}
Javadi,~S.; R{\o}yne,~A. {Adhesive forces between two cleaved calcite surfaces
  in NaCl solutions: The importance of ionic strength and normal loading}.
  \emph{Journal of Colloid and Interface Science} \textbf{2018}, \emph{532},
  605--613\relax
\mciteBstWouldAddEndPuncttrue
\mciteSetBstMidEndSepPunct{\mcitedefaultmidpunct}
{\mcitedefaultendpunct}{\mcitedefaultseppunct}\relax
\EndOfBibitem
\bibitem[Dziadkowiec \latin{et~al.}(2019)Dziadkowiec, Zareeipolgardani, Dysthe,
  and R{\o}yne]{Dziadkowiec2019}
Dziadkowiec,~J.; Zareeipolgardani,~B.; Dysthe,~D.~K.; R{\o}yne,~A. {Nucleation
  in confinement generates long-range repulsion between rough calcite
  surfaces}. \emph{Scientific Reports} \textbf{2019}, \emph{9}, 1--15\relax
\mciteBstWouldAddEndPuncttrue
\mciteSetBstMidEndSepPunct{\mcitedefaultmidpunct}
{\mcitedefaultendpunct}{\mcitedefaultseppunct}\relax
\EndOfBibitem
\bibitem[Brunsteiner and Price(2004)Brunsteiner, and Price]{Brunsteiner2004}
Brunsteiner,~M.; Price,~S.~L. {Surface structure of a complex inorganic crystal
  in aqueous solution from classical molecular simulation}. \emph{Journal of
  Physical Chemistry B} \textbf{2004}, \emph{108}, 12537--12546\relax
\mciteBstWouldAddEndPuncttrue
\mciteSetBstMidEndSepPunct{\mcitedefaultmidpunct}
{\mcitedefaultendpunct}{\mcitedefaultseppunct}\relax
\EndOfBibitem
\bibitem[Brunsteiner \latin{et~al.}(2005)Brunsteiner, Jones, Pratola, Price,
  and Simons]{Brunsteiner2005}
Brunsteiner,~M.; Jones,~A.~G.; Pratola,~F.; Price,~S.~L.; Simons,~S.~J. {Toward
  a molecular understanding of crystal agglomeration}. 2005;
  \url{https://pubs.acs.org/sharingguidelines}\relax
\mciteBstWouldAddEndPuncttrue
\mciteSetBstMidEndSepPunct{\mcitedefaultmidpunct}
{\mcitedefaultendpunct}{\mcitedefaultseppunct}\relax
\EndOfBibitem
\bibitem[Renard \latin{et~al.}(2012)Renard, Beaupr{\^{c}}tre, Voisin, Zigone,
  Candela, Dysthe, and Gratier]{Renard2012}
Renard,~F.; Beaupr{\^{c}}tre,~S.; Voisin,~C.; Zigone,~D.; Candela,~T.;
  Dysthe,~D.~K.; Gratier,~J.~P. {Strength evolution of a reactive frictional
  interface is controlled by the dynamics of contacts and chemical effects}.
  \emph{Earth and Planetary Science Letters} \textbf{2012}, \emph{341-344},
  20--34\relax
\mciteBstWouldAddEndPuncttrue
\mciteSetBstMidEndSepPunct{\mcitedefaultmidpunct}
{\mcitedefaultendpunct}{\mcitedefaultseppunct}\relax
\EndOfBibitem
\bibitem[Anders \latin{et~al.}(2014)Anders, Laubach, and Scholz]{Anders2014}
Anders,~M.~H.; Laubach,~S.~E.; Scholz,~C.~H. {Microfractures: A review}.
  \emph{Journal of Structural Geology} \textbf{2014}, \emph{69}, 377--394\relax
\mciteBstWouldAddEndPuncttrue
\mciteSetBstMidEndSepPunct{\mcitedefaultmidpunct}
{\mcitedefaultendpunct}{\mcitedefaultseppunct}\relax
\EndOfBibitem
\bibitem[Rodr{\'{i}}guez-S{\'{a}}nchez
  \latin{et~al.}(2020)Rodr{\'{i}}guez-S{\'{a}}nchez, Liberto, Barentin, and
  Dysthe]{Rodriguez-Sanchez2020}
Rodr{\'{i}}guez-S{\'{a}}nchez,~J.; Liberto,~T.; Barentin,~C.; Dysthe,~D.
  {Mechanisms of phase transformation and creating mechanical strength in a
  sustainable calcium carbonate cement}. \emph{Preprints} \textbf{2020}, \relax
\mciteBstWouldAddEndPunctfalse
\mciteSetBstMidEndSepPunct{\mcitedefaultmidpunct}
{}{\mcitedefaultseppunct}\relax
\EndOfBibitem
\bibitem[Niemeijer \latin{et~al.}(2008)Niemeijer, Marone, and
  Elsworth]{Niemeijer2008}
Niemeijer,~A.; Marone,~C.; Elsworth,~D. {Healing of simulated fault gouges
  aided by pressure solution: Results from rock analogue experiments}.
  \emph{Journal of Geophysical Research: Solid Earth} \textbf{2008},
  \emph{113}, 1--15\relax
\mciteBstWouldAddEndPuncttrue
\mciteSetBstMidEndSepPunct{\mcitedefaultmidpunct}
{\mcitedefaultendpunct}{\mcitedefaultseppunct}\relax
\EndOfBibitem
\bibitem[H{\o}gberget \latin{et~al.}(2016)H{\o}gberget, R{\o}yne, Dysthe, and
  Jettestuen]{Hogberget2016}
H{\o}gberget,~J.; R{\o}yne,~A.; Dysthe,~D.; Jettestuen,~E. {Microscopic
  modeling of confined crystal growth and dissolution}. \emph{Physical Review
  E} \textbf{2016}, \emph{94}\relax
\mciteBstWouldAddEndPuncttrue
\mciteSetBstMidEndSepPunct{\mcitedefaultmidpunct}
{\mcitedefaultendpunct}{\mcitedefaultseppunct}\relax
\EndOfBibitem
\bibitem[H{\o}gberget \latin{et~al.}(2020)H{\o}gberget, Dysthe, and
  Jettestuen]{Hogberget2020b}
H{\o}gberget,~J.; Dysthe,~D.~K.; Jettestuen,~E. {Direct Coupling of Free
  Diffusion Models to Microscopic Models of Confined Crystal Growth and
  Dissolution}. \emph{arXiv:cond-mat} \textbf{2020}, \emph{2006.01433},
  1--13\relax
\mciteBstWouldAddEndPuncttrue
\mciteSetBstMidEndSepPunct{\mcitedefaultmidpunct}
{\mcitedefaultendpunct}{\mcitedefaultseppunct}\relax
\EndOfBibitem
\bibitem[Mutisya \latin{et~al.}(2017)Mutisya, Kirch, {De Almeida},
  S{\'{a}}nchez, and Miranda]{Mutisya2017}
Mutisya,~S.~M.; Kirch,~A.; {De Almeida},~J.~M.; S{\'{a}}nchez,~V.~M.;
  Miranda,~C.~R. {Molecular Dynamics Simulations of Water Confined in Calcite
  Slit Pores: An NMR Spin Relaxation and Hydrogen Bond Analysis}. \emph{Journal
  of Physical Chemistry C} \textbf{2017}, \emph{121}, 6674--6684\relax
\mciteBstWouldAddEndPuncttrue
\mciteSetBstMidEndSepPunct{\mcitedefaultmidpunct}
{\mcitedefaultendpunct}{\mcitedefaultseppunct}\relax
\EndOfBibitem
\bibitem[Collin \latin{et~al.}(2018)Collin, Gin, Dazas, Mahadevan, Du, and
  Bourg]{Collin2018}
Collin,~M.; Gin,~S.; Dazas,~B.; Mahadevan,~T.; Du,~J.; Bourg,~I.~C. {Molecular
  Dynamics Simulations of Water Structure and Diffusion in a 1 nm Diameter
  Silica Nanopore as a Function of Surface Charge and Alkali Metal Counterion
  Identity}. \emph{Journal of Physical Chemistry C} \textbf{2018}, \emph{122},
  17764--17776\relax
\mciteBstWouldAddEndPuncttrue
\mciteSetBstMidEndSepPunct{\mcitedefaultmidpunct}
{\mcitedefaultendpunct}{\mcitedefaultseppunct}\relax
\EndOfBibitem
\bibitem[Benz \latin{et~al.}(2006)Benz, Rosenberg, Kramer, and
  Israelachvili]{benz2006deformation}
Benz,~M.; Rosenberg,~K.~J.; Kramer,~E.~J.; Israelachvili,~J.~N. The deformation
  and adhesion of randomly rough and patterned surfaces. \emph{The Journal of
  Physical Chemistry B} \textbf{2006}, \emph{110}, 11884--11893\relax
\mciteBstWouldAddEndPuncttrue
\mciteSetBstMidEndSepPunct{\mcitedefaultmidpunct}
{\mcitedefaultendpunct}{\mcitedefaultseppunct}\relax
\EndOfBibitem
\end{mcitethebibliography}

\end{document}